%% file: 0-main.tex
\documentclass{article}

\usepackage[preprint]{neurips_2026}

\usepackage[utf8]{inputenc} 
\usepackage[T1]{fontenc}    
\usepackage{hyperref}       
\usepackage{url}            
\usepackage{booktabs}       
\usepackage{amsfonts}       
\usepackage{nicefrac}       
\usepackage{microtype}      
\usepackage{xcolor}         
\usepackage{booktabs}
\usepackage{tabularx}
\usepackage{array}

\usepackage{xspace}
\usepackage{graphicx}
\usepackage{subcaption}
\usepackage{amsmath}
\usepackage[normalem]{ulem}
\usepackage{enumitem}

\newcommand{\sysname}{\textit{NPUsper}\xspace}

\title{\sysname: Eliminating Redundant Computation for Real-Time Whisper on Mobile NPUs}

\author{%
\textbf{Sihyeon Lee}$^{1}$ \quad
\textbf{Hojeong Lee}$^{2}$ \quad
\textbf{Sungwon Woo}$^{1}$ \\
\textbf{Chengpo Yan}$^{2}$ \quad
\textbf{Suman Banerjee}$^{2,\dagger}$ \quad
\textbf{Seyeon Kim}$^{1,\dagger}$ \\
\normalfont
$^{1}$Korea University \quad
$^{2}$University of Wisconsin--Madison \\
\texttt{xl6047@korea.ac.kr} \quad
\texttt{hojeong.lee@wisc.edu} \quad
\texttt{vpfkfl753@gmail.com} \\
\texttt{cyan46@wisc.edu} \quad
\texttt{suman@cs.wisc.edu} \quad
\texttt{seyeon625@korea.ac.kr}
}

\begin{document}
\maketitle

\begingroup
\renewcommand\thefootnote{}
\footnotetext{%
$^{\dagger}$Corresponding authors.
}
\endgroup

\begin{abstract}
We present \sysname\footnote{\sysname combines ``NPU'' and ``Whisper,'' with ``NP'' also referring to no padding to avoid padding overhead.}, a live transcription system that makes Whisper efficient on mobile NPUs by eliminating redundant computation. To avoid the heavy padding used by prior streaming systems, \sysname detects hallucinated tokens online from temporal patterns in decoder cross-attention, allowing each inference round to process short audio inputs with minimal carryover. For efficient mobile-NPU execution, we propose controlled unrolling, which executes autoregressive decoding as \(K\)-step chunk graphs, removing unnecessary KV-cache computation and reducing graph-dispatch overhead. \sysname achieves up to \textbf{4.84$\times$} lower per-word latency, up to \textbf{33.2$\times$} lower time-to-first-token (TTFT), and up to \textbf{88.64\%} lower average power consumption compared with baselines, while maintaining comparable transcription accuracy. The code is available at \url{https://github.com/npusper/NPUsper}.

\end{abstract}

\input{1-introduction}
\input{2-background}

\input{3-motivation}
\input{4-design}
\input{5-results}
\input{6-conclusion}

\bibliographystyle{unsrtnat}
\bibliography{references}

\newpage

\appendix
\input{7-appendix}

\end{document}

%% file: 1-introduction.tex
\section{Introduction}
On-device AI inference is becoming increasingly important for mobile applications due to its advantages in privacy, latency, and reliability. Among these applications, real-time speech transcription has emerged as a key capability for intelligent assistants and accessibility services, where timely and accurate transcription is critical to user experience~\citep{apple_siri, labiausse2025highfidelity}. Enabling such functionality directly on mobile devices is particularly desirable, as it avoids the latency and connectivity limitations of cloud-based processing while preserving user privacy.

Recent systems increasingly rely on large foundation models for robust transcription, and Whisper~\citep{radford2023robust} has emerged as a \textit{de facto} standard due to its strong accuracy and scalability. As a result, there is growing interest in running Whisper directly on mobile devices for real-time, on-device transcription~\citep{machavcek2023turning, wang2025whisperflow, wang2024simul, machavcek2025simultaneous}. To meet the stringent latency and energy requirements of real-time processing, modern mobile platforms are equipped with Neural Processing Units (NPUs) that offer high-throughput and energy-efficient inference. However, fully leveraging mobile NPUs for real-time Whisper remains challenging, due to fundamental mismatches between Whisper's streaming inference pipeline and the execution characteristics of NPUs.

\begin{figure}[t]
    \centering

    \begin{subfigure}[t]{0.533\linewidth}
        \centering
        \includegraphics[width=\linewidth]{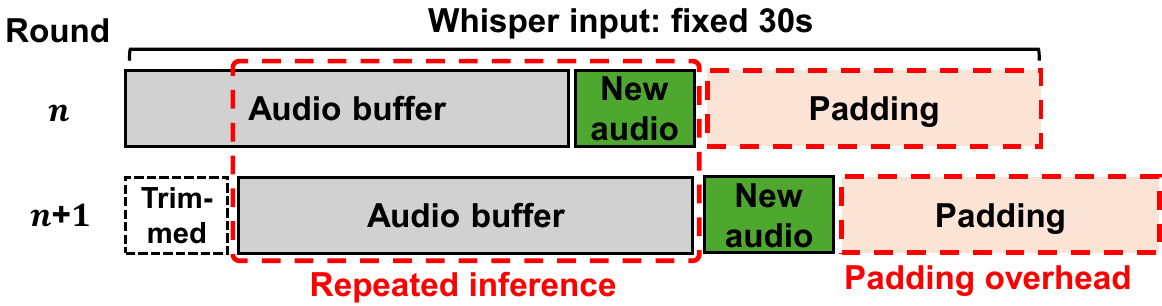}
        \caption{Whisper-Streaming, SimulStreaming, and Simul-Whisper}
        \label{fig:buffer-others}
    \end{subfigure}
    \hfill
    \begin{subfigure}[t]{0.285\linewidth}
        \centering
        \includegraphics[width=\linewidth]{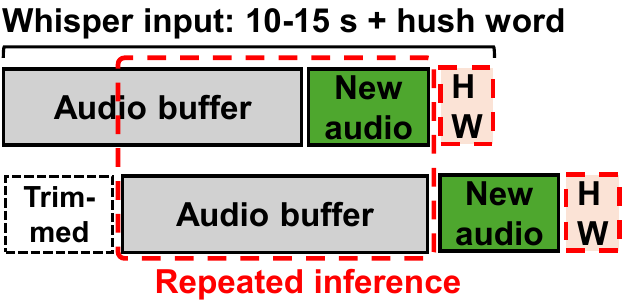}
        \caption{WhisperFlow}
        \label{fig:buffer-whisperflow}
    \end{subfigure}
    \hfill
    \begin{subfigure}[t]{0.152\linewidth}
        \centering
        \includegraphics[width=\linewidth]{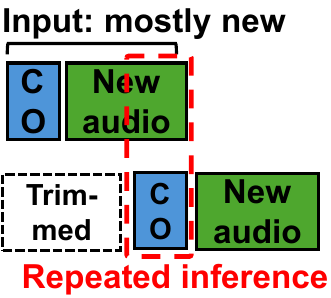}
        \caption{Ours}
        \label{fig:buffer-ours}
    \end{subfigure}

    \caption{Redundant computation in existing systems. Existing systems construct each Whisper input by buffering previously processed audio with newly arrived audio and padding it to a fixed 30-second window to mitigate hallucinations. WhisperFlow~\citep{wang2025whisperflow} reduces padding overhead by replacing long padding with a trained short audio segment, called a hush word, but still retains substantial overlap between consecutive inference inputs. HW denotes hush word and CO denotes audio carryover.}
    \label{fig:buffer-comparison}
\end{figure}

First, existing streaming approaches for Whisper incur substantial redundant computation. Whisper is prone to hallucinations when processing short audio inputs that deviate from its 30-second training distribution~\citep{wang2025whisperflow}. To mitigate this, prior systems rely on padding and overlapping audio buffers, effectively reprocessing previously seen inputs to stabilize decoding as illustrated in Figure~\ref{fig:buffer-comparison}. However, this design treats padding as a necessity, even though its primary role is to suppress hallucinations rather than to improve acoustic coverage. As a result, a large fraction of computation is spent on redundant or artificial inputs, while newly arrived speech accounts for only a small portion of each inference step.

Second, Whisper's autoregressive decoding is poorly matched to the static execution model of mobile NPUs. NPUs rely on precompiled static graphs with fixed tensor shapes, whereas decoding involves dynamically growing key-value (KV) caches. Existing approaches either execute attention over the maximum sequence length at every decoding step, incurring substantial redundant computation, or rely on step-specific graphs, which introduce significant runtime overhead due to frequent graph dispatch.

To address these challenges, we present \sysname, a system for real-time on-device Whisper transcription on mobile NPUs. 
Our key insight is that the inefficiency of streaming Whisper stems from two structural mismatches:
\textbf{(i)} hallucination-induced padding overhead caused by unstable temporal grounding under short inputs, and
\textbf{(ii)} the mismatch between dynamic autoregressive decoding and static-graph NPU execution.
\sysname resolves these mismatches through temporal-grounding-aware decoding control and NPU-aware controlled unrolling.

We implement and evaluate \sysname on commercial mobile devices \citep{samsung_galaxy_s25, qualcomm_laptop_products} and show that it achieves up to \textbf{4.84$\times$} lower per-word latency, up to \textbf{33.2$\times$} lower time-to-first-token (TTFT), and up to \textbf{88.64\%} lower average power consumption compared with baselines, while maintaining comparable transcription accuracy. 

%% file: 2-background.tex
\section{Background and Related Work}

\subsection{Live Transcription Systems with Whisper}

\textbf{Whisper.} Whisper~\citep{radford2023robust} is an encoder--decoder Transformer model widely used for speech transcription due to its strong accuracy and robustness. Although originally designed for offline inference, it has been increasingly adopted in live transcription systems through buffering and specialized decoding strategies~\citep{machavcek2023turning, wang2025whisperflow, wang2024simul, machavcek2025simultaneous}.

\textbf{Live transcription systems.} Processing continuously arriving speech in real time requires balancing low latency with transcription accuracy~\citep{yu2016automatic}. To achieve this, existing systems repeatedly invoke Whisper on partial inputs and rely on decoding policies such as Local Agreement~\citep{liu2020low} and AlignAtt~\citep{papi2023alignatt} to determine when partial hypotheses are stable enough to emit.

\subsection{Neural Processing Units}

Neural processing units (NPUs) execute models using precompiled static graphs with fixed tensor shapes, enabling high efficiency~\citep{samsung_galaxy_s25, meta_quest, qualcomm_laptop_products} but limiting support for dynamic workloads such as autoregressive decoding.

\textbf{Static-graph execution.} 
To exploit these hardware characteristics, NPUs typically execute models using precompiled static graphs, where tensor shapes, memory layouts, and execution schedules are fixed. This enables efficient mapping of operations onto parallel compute units, improves data reuse through on-chip SRAM buffering, and reduces costly memory movement. As a result, static-graph execution allows NPUs to achieve high utilization of compute resources while minimizing runtime overhead and energy consumption~\citep{lee2021architecture, hameed2010understanding}.

\textbf{Implications for live transcription.} 
This execution model is well-suited to workloads with fixed or predictable tensor shapes, such as vision models with static input resolutions~\citep{jung2025aria, tan2021efficient}. In contrast, live transcription involves continuously arriving audio inputs and incremental decoding, which naturally introduce variability in input sizes and execution patterns~\citep{yu2016automatic}. As a result, efficiently mapping such workloads onto static-graph execution often requires structured input handling or approximation strategies that align dynamic inputs with a limited set of predefined execution configurations~\citep{khomenko2016accelerating}.

%% file: 3-motivation.tex
\section{Motivation}\label{sec:motivation}

\begin{figure}[t]
    \centering

    \begin{minipage}[t]{0.49\linewidth}
        \centering

        \begin{minipage}[t]{0.49\linewidth}
            \centering
            \includegraphics[width=\linewidth]{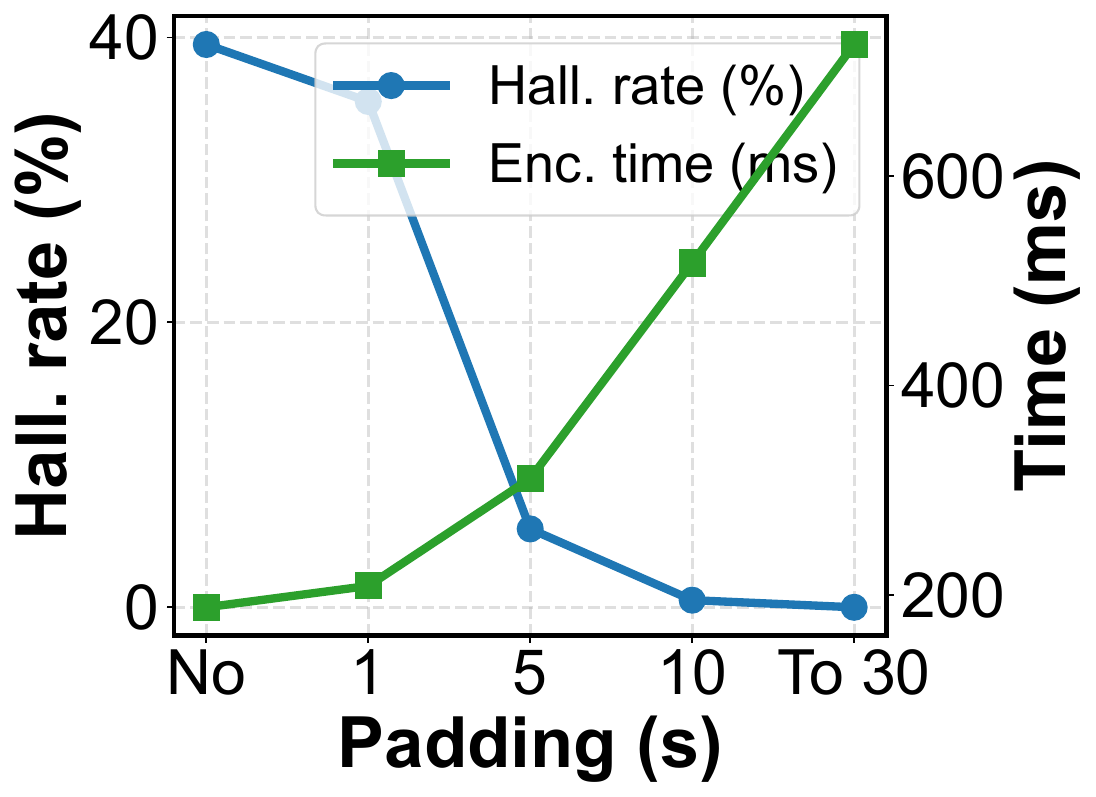}
            {\small (a) Padding vs. hallucination/overhead}
            \label{fig:padding_hallucination}
        \end{minipage}
        \hfill
        \begin{minipage}[t]{0.49\linewidth}
            \centering
            \includegraphics[width=\linewidth]{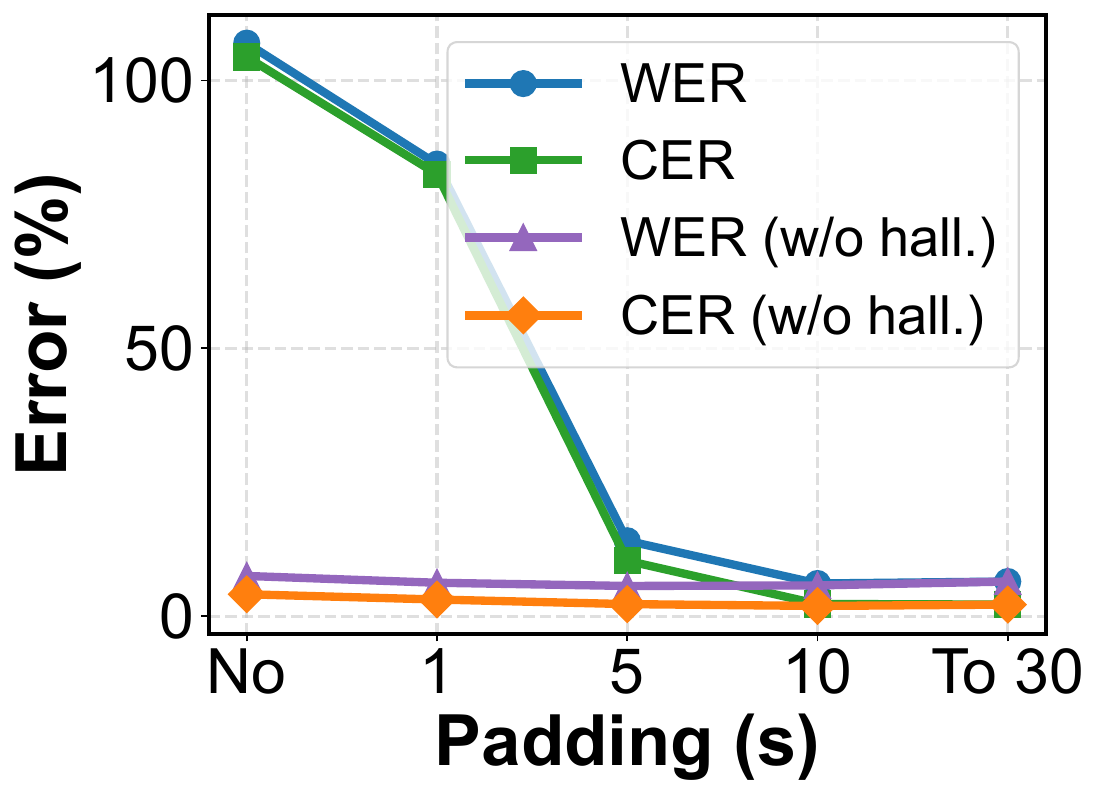}
            {\small (b) Padding vs. accuracy}
            \label{fig:padding_error}
        \end{minipage}

        \captionof{figure}{Padding reduces hallucination and improves accuracy, but increases processing overhead.}
        \label{fig:motivation-padding_tradeoff}
    \end{minipage}
    \hfill
    \begin{minipage}[t]{0.49\linewidth}
        \centering

        \begin{minipage}[t]{0.49\linewidth}
            \centering
            \includegraphics[width=\linewidth]{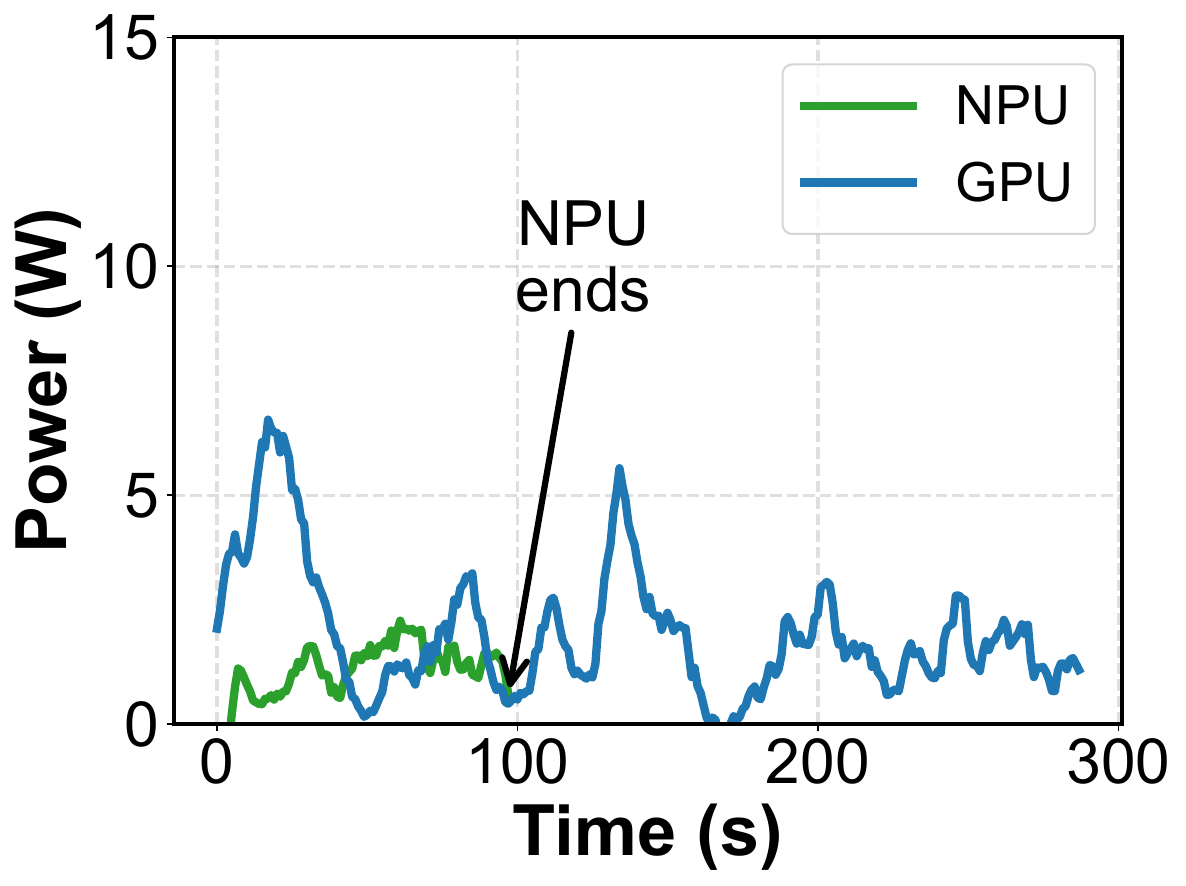}
            {\small (a) Power and throughput}
            \label{fig:npu-efficiency}
        \end{minipage}
        \hfill
        \begin{minipage}[t]{0.49\linewidth}
            \centering
            \includegraphics[width=\linewidth]{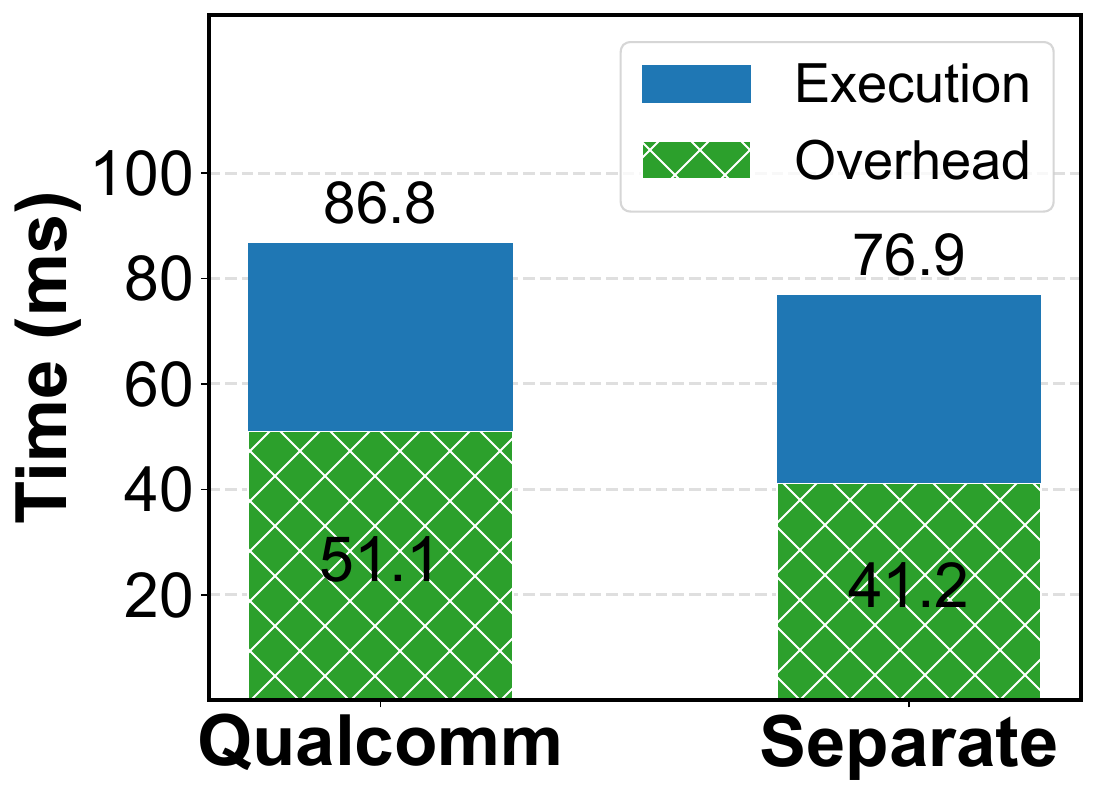}
            {\small (b) Decoding latency breakdown on NPU}
            \label{fig:npu-static-graph}
        \end{minipage}

        \captionof{figure}{NPUs provide faster and more energy-efficient inference, while static-graph execution inefficiency motivates NPU-aware decoding.}
        \label{fig:motivation-npu}
    \end{minipage}

\end{figure}

\subsection{Computational Cost of Padding and Hallucination Risk} \label{sec:motivation-padding}

\textbf{Whisper hallucination.} Whisper exhibits hallucination behavior, generating text that is not grounded in the input audio \citep{kalai2025language, baranski2025investigation, koenecke2024careless}. In speech transcription, this typically manifests as repeated words or phrases \citep{koenecke2024careless, wang2025whisperflow}. 
As shown in Figure~\ref{fig:motivation-padding_tradeoff}(a), hallucinations become more frequent when processing short audio inputs without padding. To mitigate this, existing systems extend the input with padding toward the 30-second training window, which stabilizes decoding. Correspondingly, Figure~\ref{fig:motivation-padding_tradeoff}(b) shows that the word error rate (WER) and character error rate (CER) decrease as the input length approaches 30 seconds. This behavior stems from Whisper being trained on fixed 30-second audio segments \citep{radford2023robust, wang2025whisperflow}, indicating that padding primarily serves to suppress hallucinations rather than to provide additional acoustic context.

\textbf{Padding overhead.} While padding mitigates hallucinations and improves WER/CER, it incurs significant computational overhead, especially on resource-constrained mobile devices. As shown in Figure \ref{fig:motivation-padding_tradeoff}(a), the encoder processing time increases substantially as the padded input length approaches 30 seconds, reaching around 720 ms, which is 3.8$\times$ higher than that without padding. 
The smaller gap observed in Figure~\ref{fig:motivation-padding_tradeoff}(a) is due to the longer input duration in this experiment (approximately 9 seconds), which exceeds the shorter segments typically used in live transcription. We focus on encoding latency in this motivation study because decoding latency depends on both input length and the number of generated tokens, making it difficult to isolate the effect of padding.

\textbf{Opportunity.} As shown in Figure~\ref{fig:motivation-padding_tradeoff}(b), when hallucinated tokens generated beyond the reference transcription are removed in offline post-processing (w/o hall. in the figure), Whisper inference without padding achieves WER and CER close to those obtained with full 30-second padding. 
These observations suggest that hallucinations are fundamentally associated with failures in temporal grounding rather than insufficient acoustic context itself.
Therefore, if temporally ungrounded tokens can be detected online during decoding, it becomes possible to eliminate padding overhead while preserving transcription accuracy.
To realize this opportunity, we analyze cross-attention patterns to detect hallucinations and avoid unnecessary computation in \sysname, detailed in \S~\ref{sec:hallucination-detection}.

\subsection{Dynamic Speech Processing on Static-graph NPUs} \label{motivation-static}
\textbf{NPU execution mismatch.} NPUs offer high inference throughput and energy efficiency, making them attractive for on-device live transcription systems. As shown in Figure~\ref{fig:motivation-npu}(a), on the same 200 Whisper inference tasks, the NPU completes inference 3$\times$ faster than the GPU and reduces average power consumption from 1.99~W to 1.14~W. However, using NPUs for live transcription introduces an execution mismatch: input length and autoregressive decoding are dynamic, whereas NPUs rely on static graphs with fixed shapes and control flow. While variability in input length can be handled relatively easily using bucketing \citep{khomenko2016accelerating}, autoregressive decoding is more challenging to execute efficiently on NPUs. Specifically, the KV cache grows with each generated token, creating a mismatch between dynamic autoregressive decoding and static-graph execution on NPUs.

\textbf{KV-cache inefficiency.} To enable autoregressive decoding on NPUs, Qualcomm adopts a design that fixes the encoder input to 30 seconds and allocates the KV cache for the maximum decoded sequence length, $M=200$~\citep{qualcomm_whisper_base}. We focus on the decoding-side inefficiency, as the encoding-side overhead from input padding was discussed in \S~\ref{sec:motivation-padding}. In autoregressive decoding, the KV cache grows incrementally with the number of generated tokens, so early decoding steps require only minimal attention computation. Under Qualcomm's design, however, the NPU must execute attention over all $M=200$ positions at every decoding step, even when the actual sequence length is much shorter. As a result, substantial redundant computation is incurred throughout decoding.

\textbf{Runtime overhead.} A straightforward way to reduce KV-cache inefficiency in Qualcomm's fixed-shape design is to compile separate step-specific graphs that match the current KV-cache length. However, this requires dispatching a different graph at each token-generation step, causing substantial runtime overhead (Figure~\ref{fig:motivation-npu}(b)). Another alternative is to unroll $n$ decoding steps into a single graph to amortize this overhead. Once dispatched, however, the graph must execute all $n$ steps, even if EOS or a stop condition occurs earlier. This highlights the need for a balanced NPU decoding design that reduces redundant KV-cache computation, minimizes runtime overhead, and preserves early termination (\S~\ref{sec:controlled-unrolling}). 

%% file: 4-design.tex
\section{\sysname System Design}
\begin{figure}[t]
    \centering
    \includegraphics[width=\linewidth]{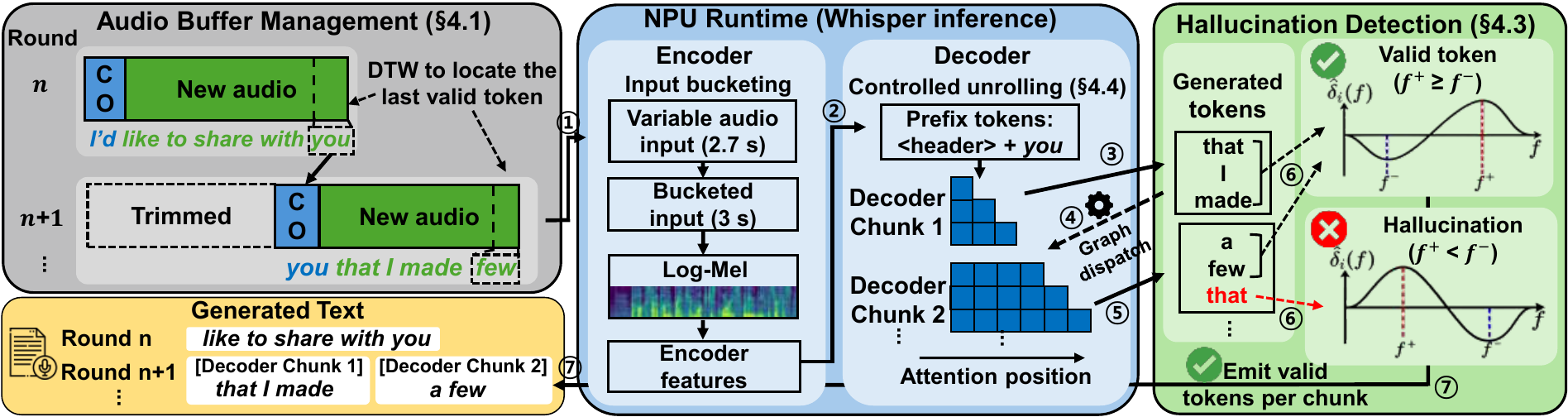}
    \caption{\sysname overview.}
    \label{fig:design-overview}
\end{figure}
Figure~\ref{fig:design-overview} presents an overview of \sysname.
At each inference round, \sysname constructs a short Whisper input from newly arrived audio and carry-over context (\S~\ref{sec:audio-buffer}).
The input is mapped to an NPU-compatible bucket, and decoding is executed with controlled unrolling (\S~\ref{sec:controlled-unrolling}).
Generated tokens are then validated using cross-attention dynamics (\S~\ref{sec:cross-attention} and \S~\ref{sec:hallucination-detection}).
Valid tokens are emitted immediately, while hallucinated tokens stop decoding and determine the carry-over state for the next round.

\subsection{Audio Buffer Management}\label{sec:audio-buffer}
The audio buffer manager maintains the incoming audio stream and constructs each input from carry-over audio and newly arrived audio. Unlike prior systems that rely on padding, \sysname constructs inputs using only real audio and validated context. 
The input is mapped to a discrete bucket for NPU execution. For autoregressive decoding, \sysname initializes the decoder with prefix tokens derived from the carry-over text. \sysname validates generated tokens using cross-attention and emits only valid tokens. 
When hallucination is detected, decoding stops and the carry-over state is updated using dynamic time warping (DTW)~\citep{sakoe1978dynamic}.

\subsection{Cross-attention Characteristics in Whisper}\label{sec:cross-attention}

\begin{figure}[t]
    \centering
    \begin{subfigure}[t]{0.24\linewidth}
        \centering
        \includegraphics[width=\linewidth]{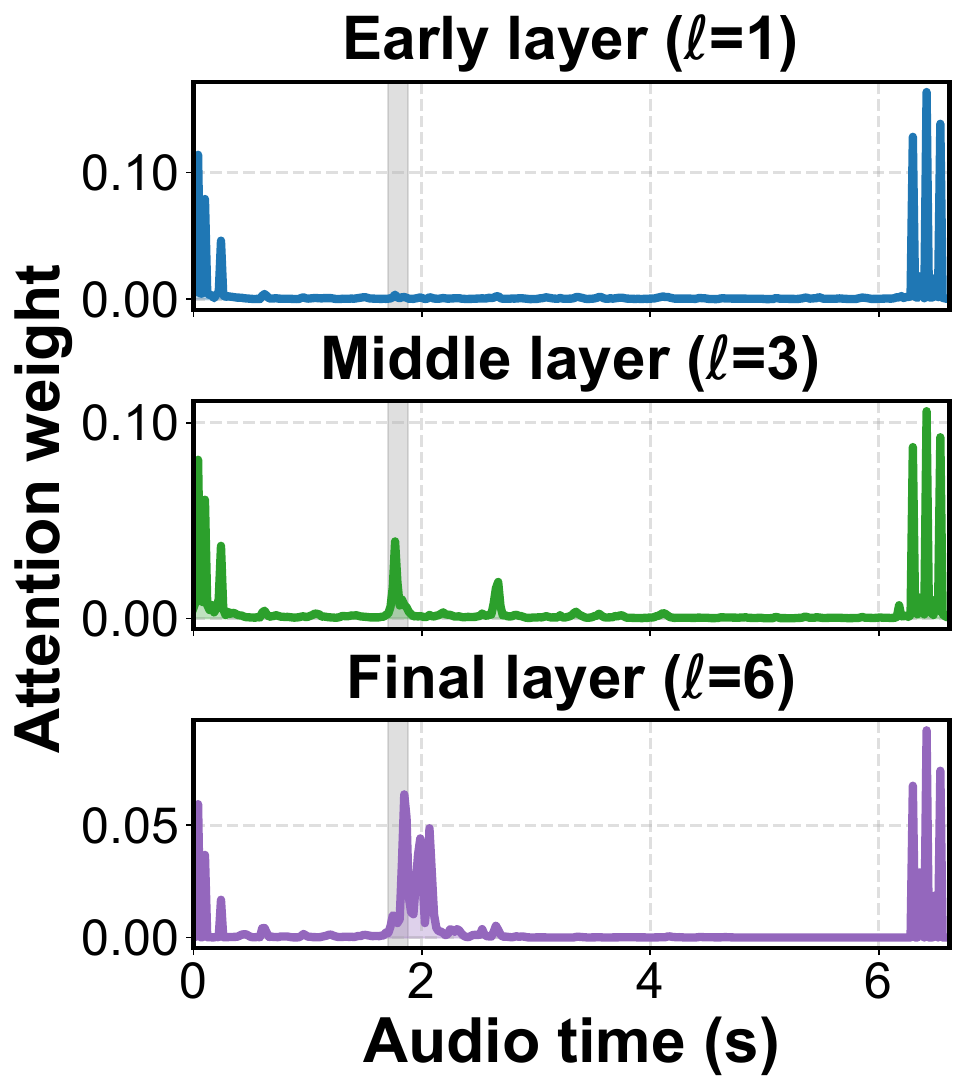}
        \caption{Token: the}
        \label{fig:token-the}
    \end{subfigure}
    \hfill
    \begin{subfigure}[t]{0.24\linewidth}
        \centering
        \includegraphics[width=\linewidth]{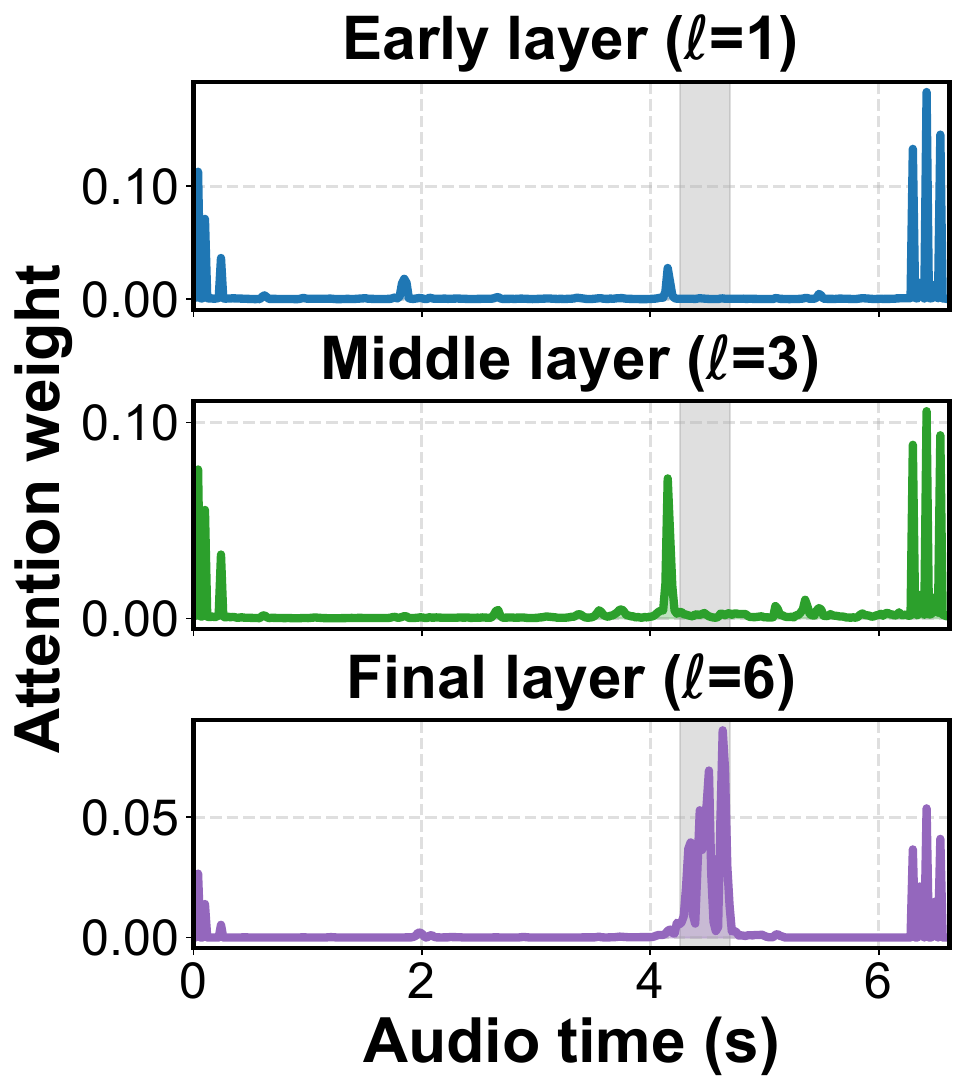}
        \caption{Token: squ}
        \label{fig:token-squ}
    \end{subfigure}
    \hfill
    \begin{subfigure}[t]{0.24\linewidth}
        \centering
        \includegraphics[width=\linewidth]{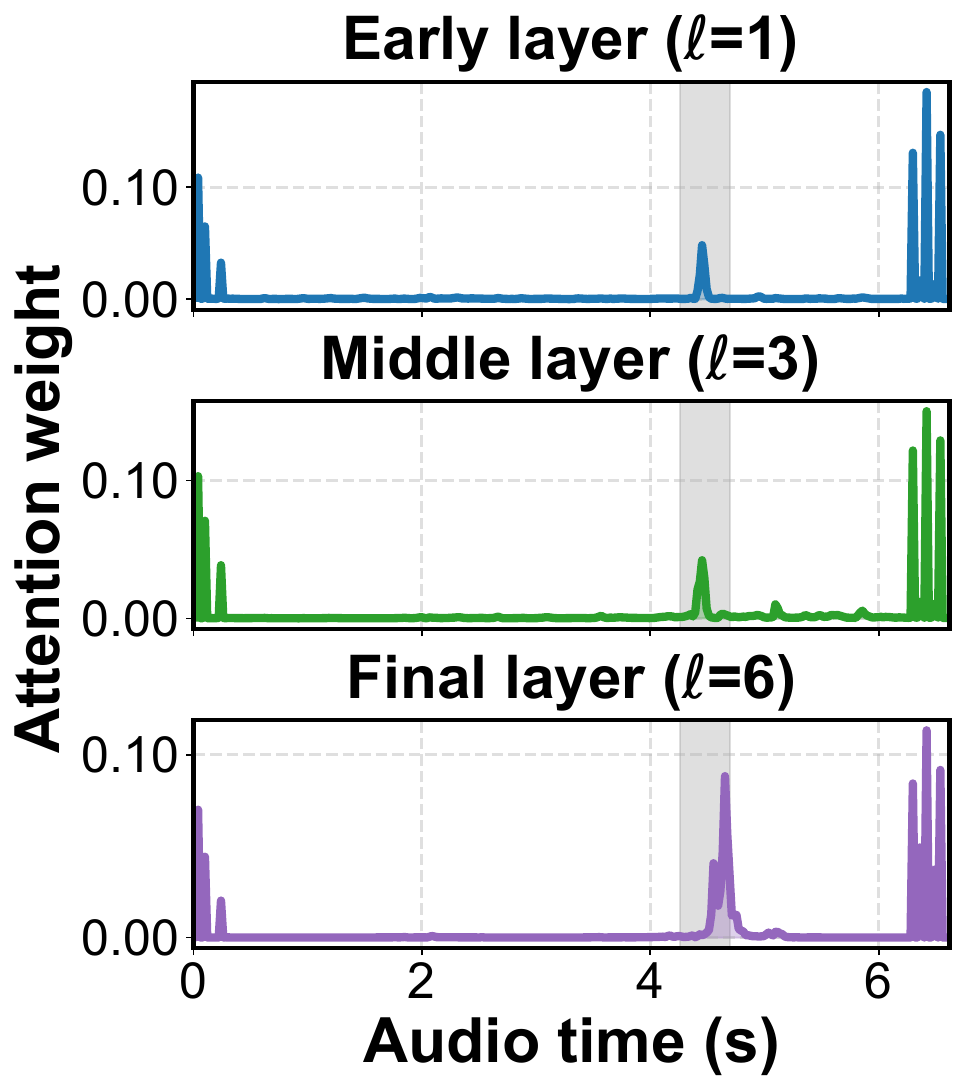}
        \caption{Token: -alid}
        \label{fig:token-alid}
    \end{subfigure}
    \hfill
    \begin{subfigure}[t]{0.24\linewidth}
        \centering
        \includegraphics[width=\linewidth]{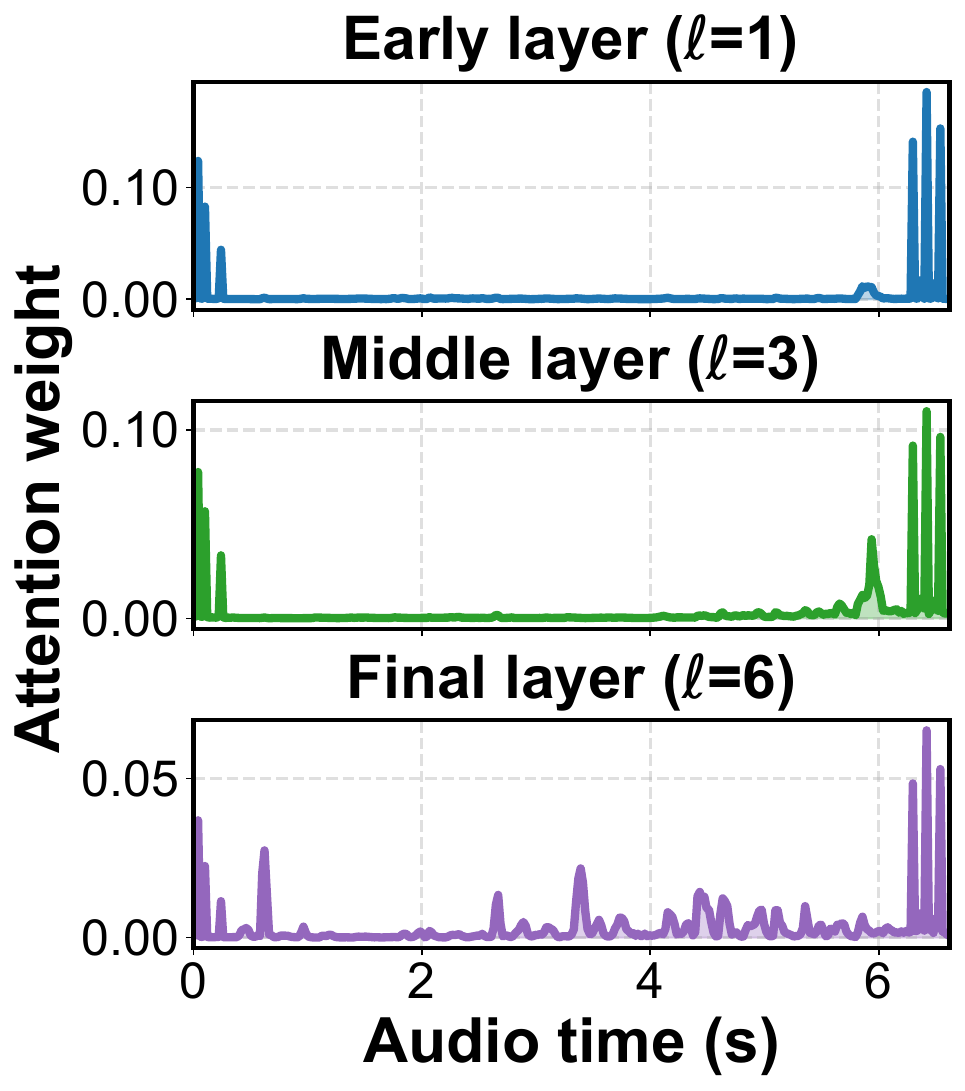}
        \caption{Token: .}
        \label{fig:token-.}
    \end{subfigure}
    \caption{Layer-wise cross-attention dynamics in Whisper over time for selected tokens from the utterance: ``After early nightfall, \uline{the} yellow lamps would light up here and there, the \uline{squalid} quarter of the brothels\uline.'' The shaded region indicates the ground-truth audio span of each token. The final decoder layer shows sharper temporal alignment than earlier layers. Content tokens form well-localized attention peaks within their spoken duration, whereas the punctuation token exhibits more diffuse cross-attention with no clear temporal anchor.}
    \label{fig:token-time-alignment}
\end{figure}

\textbf{Definition of cross-attention.} For the $i$-th generated token, we define the frame-wise cross-attention at decoder layer $l$ as the head-averaged attention:
\begin{equation}
a_i^{(l)}(f) = \frac{1}{H} \sum_{h=1}^{H} \mathrm{CrossAttn}_i^{(l,h)}(f),
\label{eq:cross-attention}
\end{equation}
where $H$ is the number of attention heads, $f$ denotes the audio frame index, and $\mathrm{CrossAttn}_i^{(l,h)}(f)$ is the softmax-normalized cross-attention weight assigned to frame $f$ at decoder layer $l$ and head $h$ when generating the $i$-th token. $a_i^{(l)}$ provides a token-level distribution over audio time, which we use to analyze temporal grounding.

\textbf{Cross-attention analysis.}
Cross-attention in Whisper indicates which audio frames contribute to each generated token. We use this signal to analyze whether tokens are temporally grounded in the input audio.
Figure~\ref{fig:token-time-alignment} shows that content tokens exhibit localized attention that follows the temporal progression of the input audio. This property is not specific to Whisper, but arises from the alignment structure of encoder-decoder speech models \citep{papi2023alignatt, papi2024streamatt}. We hypothesize that valid tokens follow a monotonic progression over audio frames, whereas hallucinated tokens violate this temporal consistency. This suggests that cross-attention can serve as a proxy for token-level grounding.
Punctuation tokens exhibit diffuse attention over the audio timeline, and subword tokens can introduce noise in temporal alignment. We therefore restrict our analysis to content tokens for reliable detection (\S~\ref{sec:hallucination-detection}).

\textbf{Layer selection.}
The temporal structure of cross-attention also varies across decoder layers. Compared with earlier layers, the final decoder layer shows clearer temporal alignment with the audio timeline (Figure~\ref{fig:token-time-alignment}), making it more suitable for tracking token-level progression. We therefore use the final layer and compare cross-attention across consecutive content tokens.

\subsection{Hallucination Detection}\label{sec:hallucination-detection}
Based on observations and insights in \S~\ref{sec:cross-attention}, we design a hallucination detection mechanism that identifies violations of temporal consistency in cross-attention patterns.

\textbf{Backward cross-attention shift.}
\sysname detects hallucinated tokens from cross-attention patterns in the final decoder layer. In Whisper, hallucinated tokens exhibit backward shifts in cross-attention over the audio timeline, rather than following the forward temporal progression of valid tokens.

\textbf{Token validation.} To operationalize this intuition, we define token-level temporal attention dynamics as follows. For each generated token $i$, we extract the head-averaged cross-attention from the final decoder layer $L$:
\begin{equation}
\tilde{a}_i(f) = a_i^{(L)}(f).
\label{eq:final-layer-cross-attn}
\end{equation}
Temporal attention changes are computed only between consecutive content tokens, excluding punctuation and subword tokens to ensure reliable temporal alignment. Let $p(i)$ denote the most recent preceding content token before token $i$. We define the frame-wise attention difference as $\delta_i(f) = \tilde{a}_i(f) - \tilde{a}_{p(i)}(f)$.
Let $\hat{\delta}_i(f)$ denote the filtered attention (Appendix~\ref{app:hallucination-detection-analysis}) difference between token $i$ and its preceding content token. 
We identify the locations of the positive and negative peaks:
\begin{equation}
f_i^{+} = \operatorname*{arg\,max}_{f} \hat{\delta}_i(f), \qquad
f_i^{-} = \operatorname*{arg\,min}_{f} \hat{\delta}_i(f).
\label{eq:peak-locations}
\end{equation}
Intuitively, $f_i^{+}$ indicates the audio frame toward which attention newly concentrates, whereas $f_i^{-}$ indicates the frame from which attention moves away. A token is considered valid if attention progresses forward along the audio timeline, i.e., $f_i^+ \geq f_i^-$. 
Otherwise, we mark it as hallucinated:
\begin{equation}
f_i^{+} < f_i^{-}.
\label{eq:attention-reversal}
\end{equation}
This condition captures violations of temporal monotonicity in cross-attention, which we treat as a signature of hallucination. This depends only on the relative ordering of attention dynamics and does not require threshold tuning.

\textbf{Decoding control.}
During decoding, \sysname stops generation when hallucination is detected or when attention reaches the end of the current audio chunk. In both cases, valid tokens are emitted, and the carry-over state is updated using DTW~\citep{sakoe1978dynamic}. This enables early termination of decoding and avoids unnecessary computation.

\subsection{Controlled Unrolling for Decoding on NPUs}\label{sec:controlled-unrolling}

\begin{figure}[t]
    \centering
    \begin{subfigure}[t]{0.27\linewidth}
        \centering
        \includegraphics[height=3.9cm]{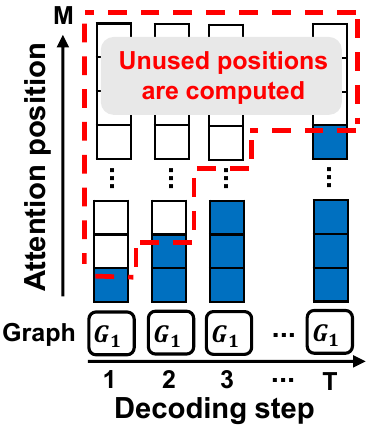}
        \caption{Fixed full-length graph}
        \label{fig:decoding-fixed}
    \end{subfigure}
    \hfill
    \begin{subfigure}[t]{0.27\linewidth}
        \centering
        \includegraphics[height=3.9cm]{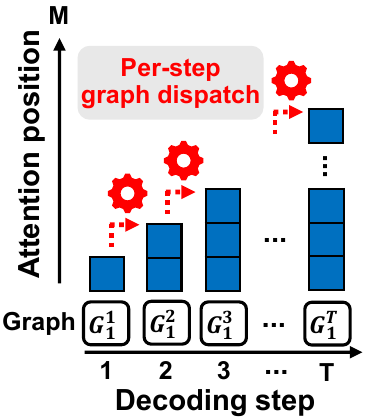}
        \caption{Step-specific graphs}
        \label{fig:decoding-separate}
    \end{subfigure}
    \hfill
    \begin{subfigure}[t]{0.4\linewidth}
        \centering
        \includegraphics[height=3.9cm]{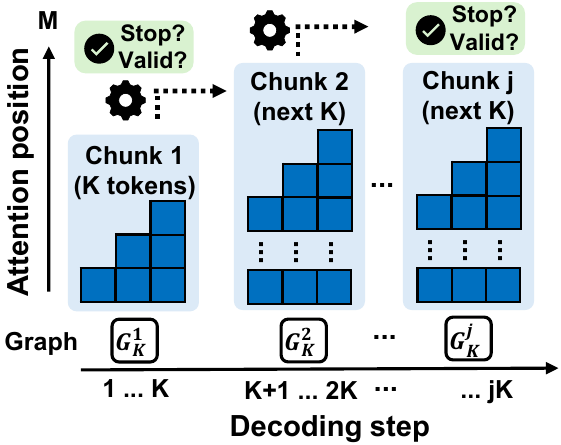}
        \caption{Controlled unrolling (Ours)}
        \label{fig:decoding-unrolling}
    \end{subfigure}

    \caption{Comparison of decoding strategies on NPUs. Fixed full-length graphs reuse the same graph $G_1$ while computing attention over the maximum KV-cache length $M$ at every step. Step-specific graphs use $G^r_1$ to match the KV-cache length required at each decoding step, but require graph dispatch for every generated token. Controlled unrolling uses $G^r_K$ to generate $K$ tokens per execution, reducing dispatch overhead while avoiding redundant KV-cache computation. Here, $G^r_q$ denotes the $r$-th graph used in a decoding strategy, where $q$ is the number of tokens generated by that graph.}
    \label{fig:design-decoding}
\end{figure}

\textbf{Controlled unrolling.}
Our system unrolls autoregressive decoding into $K$-step chunk graphs and executes them sequentially on the NPU. 
This design explicitly bridges the mismatch between dynamic autoregressive decoding and static-graph execution on NPUs.
As shown in Figure~\ref{fig:design-decoding}(\subref{fig:decoding-unrolling}), each chunk graph generates up to $K$ tokens in one NPU execution and returns the cross-attention weights defined in Equation~\eqref{eq:final-layer-cross-attn}. \sysname then applies the token validation in \S~\ref{sec:hallucination-detection} to decide whether to emit tokens, stop decoding, or dispatch the next chunk graph. This enables adaptive computation that scales with the actual decoding length rather than the maximum sequence length.

\textbf{Chunk-level decoding control.} A $K$-token chunk serves as the unit of both NPU execution and decoding control. After each chunk, \sysname applies the same decoding control conditions described in \S~\ref{sec:hallucination-detection} to decide whether to dispatch the next graph or proceed to the next inference round. Valid tokens are emitted after each chunk, rather than waiting for the entire decoding round to finish. If a hallucinated token is detected, \sysname emits only the preceding valid tokens. This chunk-level design reduces graph dispatch overhead while preserving timely token emission and hallucination handling.

\textbf{Necessary KV-cache computation.}
Controlled unrolling keeps only the KV-cache attention computation needed at each decoding step. Let $M$ be the maximum decode length of the fixed full-length graph, and let $\mathrm{MAC}(s)$ denote the number of multiply-accumulate operations for self-attention at decoding step $s$. We use $c$ to denote the per-position self-attention cost, including the query-key dot product and value aggregation across all decoder layers and attention heads. The fixed full-length graph computes attention over all $M$ positions at every step, so $\mathrm{MAC}_{\text{fixed}}(s)=cM$. In contrast, \sysname computes attention only over the $s$ positions needed at step $s$, so $\mathrm{MAC}_{\text{ours}}(s)=cs$. If decoding stops at step $T$, the resulting self-attention MAC ratio is
\begin{equation}
\frac{\sum_{s=1}^{T}\mathrm{MAC}_{\text{fixed}}(s)}
{\sum_{s=1}^{T}\mathrm{MAC}_{\text{ours}}(s)}
=
\frac{2M}{T+1}.
\label{eq:mac-ratio}
\end{equation}
This reveals a fundamental inefficiency in fixed-length decoding, where computation scales with the maximum sequence length $M$ rather than the actual decoding length $T$. 
This inefficiency becomes particularly pronounced in live transcription, where decoding often terminates after only a few tokens.
For example, with $M=200$, stopping at $T=10$ yields a $36.4\times$ reduction in self-attention MACs, while $T=20$ still yields a $19\times$ reduction.

%% file: 5-results.tex
\section{Results}\label{sec:results}

\subsection{Evaluation Setup}

We implemented and evaluated \sysname and baseline systems on two commercial mobile devices: a Samsung Galaxy S25~\citep{samsung_galaxy_s25} and a Snapdragon X Plus laptop~\citep{qualcomm_laptop_products}, hereafter referred to as S25 and X Plus, respectively. For a fair comparison, all baselines and our method run on the GPU. We additionally evaluate our method on the NPU, denoted as Ours-N, to measure the benefit of NPU execution. Appendices~\ref{app:experiment-detail} and ~\ref{app:implementation-detail} provide implementation details.

\textbf{Baselines.}
We compare \sysname with four Whisper-based live transcription systems: Whisper-Streaming (WS)~\citep{machavcek2023turning}, WhisperFlow (WF)~\citep{wang2025whisperflow}, SimulStreaming (SS)~\citep{machavcek2025simultaneous}, and Simul-Whisper (SW)~\citep{wang2024simul}.
They differ in Whisper input construction, decoder context management, and decoding policy.
WS, SS, and SW feed Whisper with 30-second sliding-window inputs, whereas WF reduces the input to roughly 15 seconds using a trained hush word.
For decoding, WS and WF use Local Agreement~\citep{liu2020low}, which repeatedly confirms partial hypotheses before emission, while SS and SW use AlignAtt~\citep{papi2023alignatt} to stop decoding based on cross-attention near chunk boundaries.
For decoder context, SS retains tokens from dropped audio segments, whereas SW discards them and uses a trained integrate-and-fire (IF) module to detect word truncation~\citep{dong2020cif}.

\textbf{Dataset.}
We evaluate on two long-form datasets, TED-LIUM 3~\citep{hernandez2018ted} and Meanwhile~\citep{radford2023robust}.
Because each end-to-end system evaluation runs in real time with the input audio, the main results use two representative TED-LIUM 3 talks, EricMead (7:39 min) and GaryFlake (5:45 min).
Additional results on the remaining TED-LIUM 3 talks and Meanwhile dataset are available in Appendix~\ref{app:additional-results}.

\textbf{Metrics.} Per-word latency ($\downarrow$) is the average latency between when a word is spoken and when it appears as text to the user~\citep{machavcek2023turning, wang2025whisperflow}. Word error rate (WER, $\downarrow$) is the fraction of incorrectly transcribed words relative to the reference transcript~\citep{radford2023robust, zhang2020transformer}. Time-to-first-token (TTFT, $\downarrow$) is the delay from the start of inference to the first text output visible to the user~\citep{kudlur2026moonshine}.

\begin{figure}[t]
    \centering
    \begin{subfigure}[t]{0.43\linewidth}
        \centering
        \includegraphics[width=\linewidth]{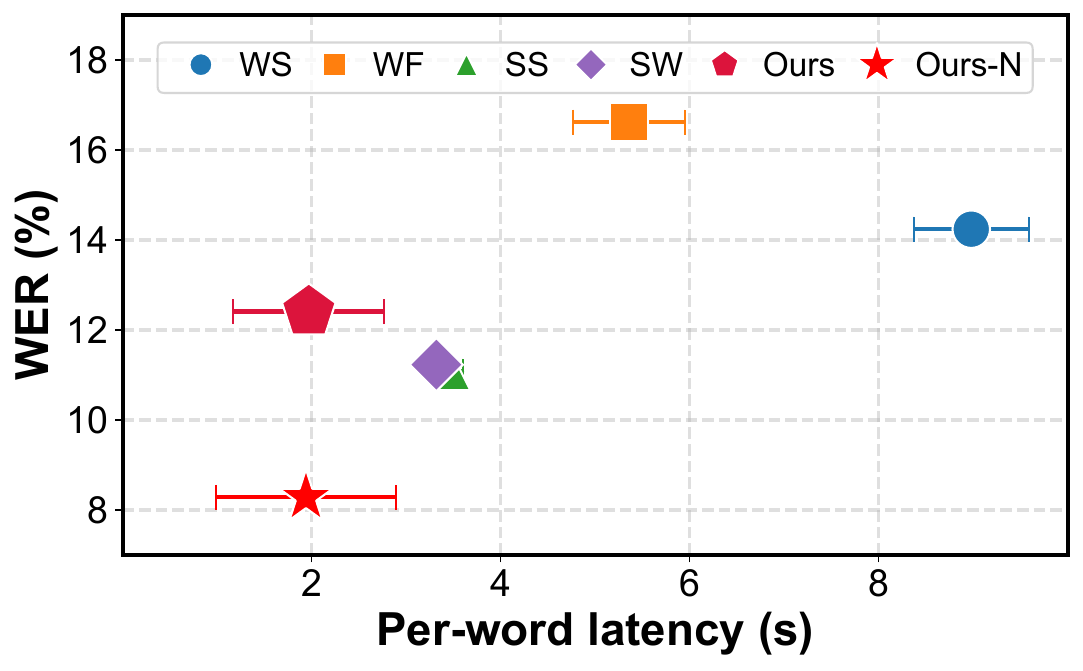}
        \caption{Samsung Galaxy S25}
        \label{fig:}
    \end{subfigure}
    \begin{subfigure}[t]{0.43\linewidth}
        \centering
        \includegraphics[width=\linewidth]{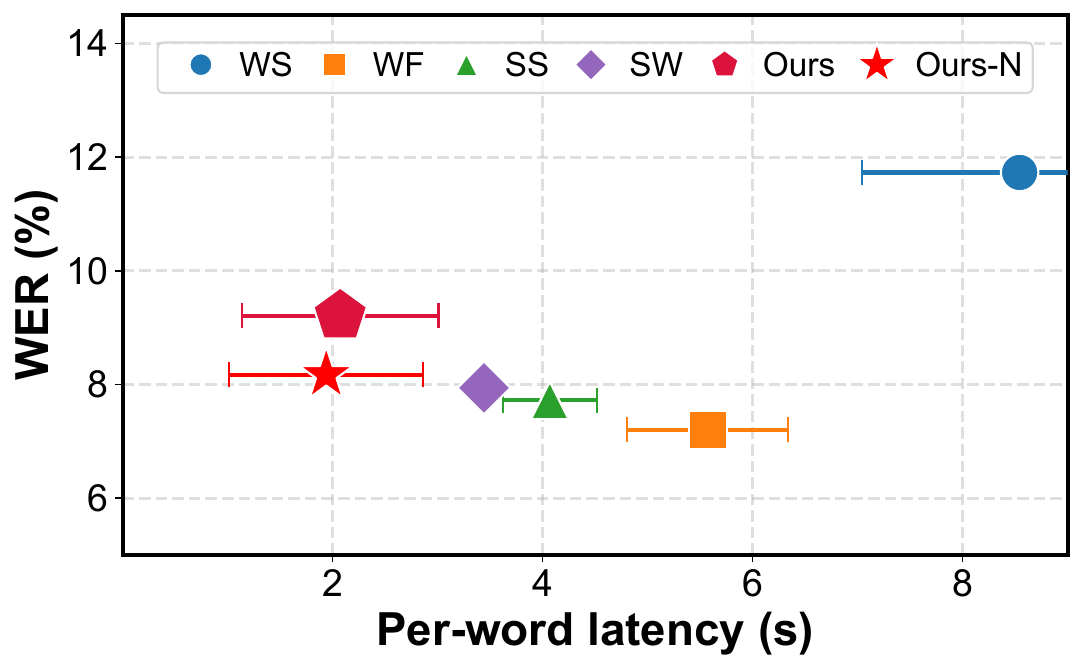}
        \caption{Snapdragon X Plus}
        \label{fig:}
    \end{subfigure}
    \caption{Per-word latency and WER comparison. \sysname achieves the lowest per-word latency on both devices with comparable transcription accuracy. Note that a 1 pp WER gap corresponds to only one additional word error per 100 reference words.}
    \label{fig:latency-vs-WER}
\end{figure}

\begin{figure}[t]
    \centering
    \begin{subfigure}[t]{0.32\linewidth}
        \centering
        \includegraphics[width=\linewidth]{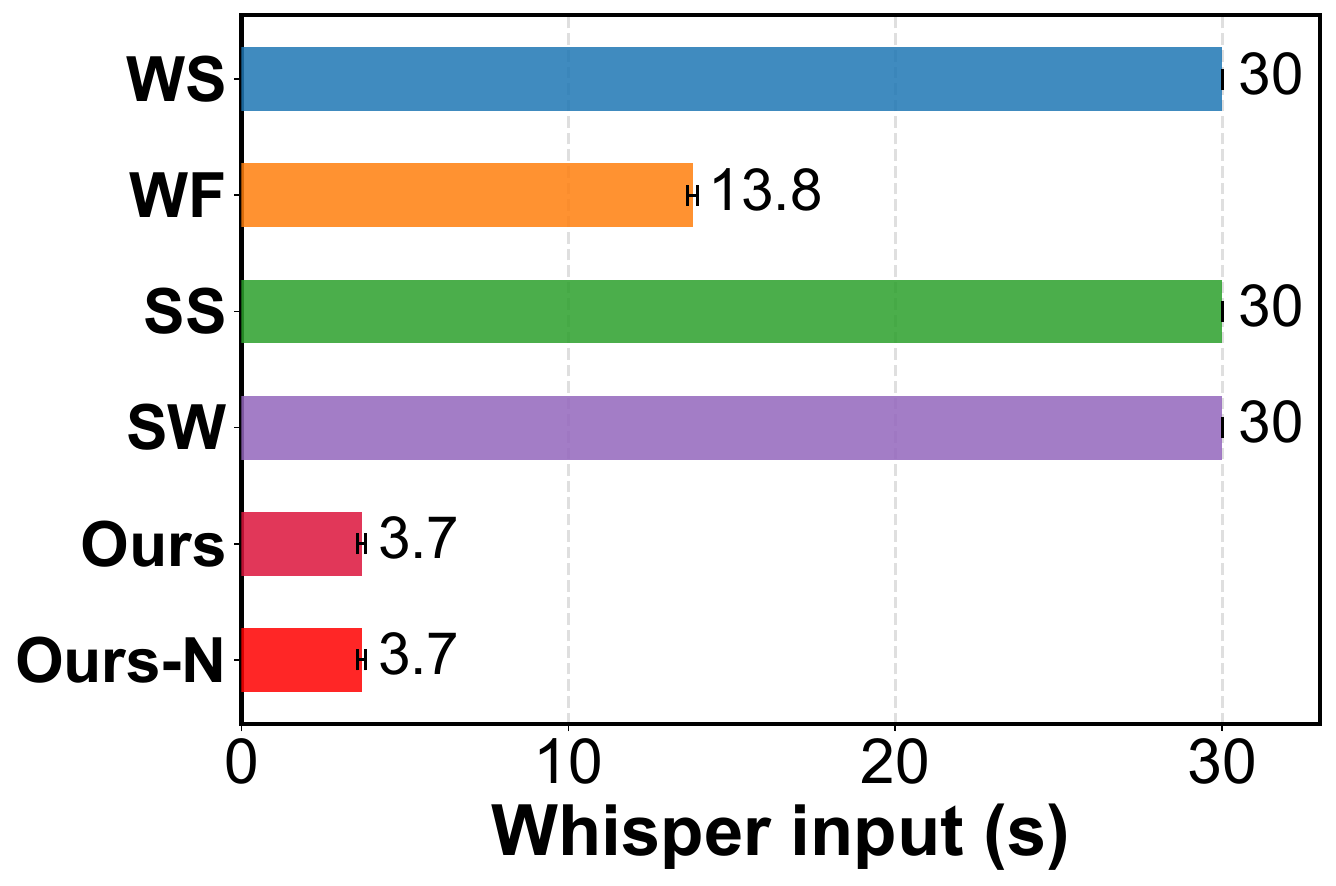}
        \caption{Average Whisper input length}
        \label{fig:}
    \end{subfigure}
    \begin{subfigure}[t]{0.32\linewidth}
        \centering
        \includegraphics[width=\linewidth]{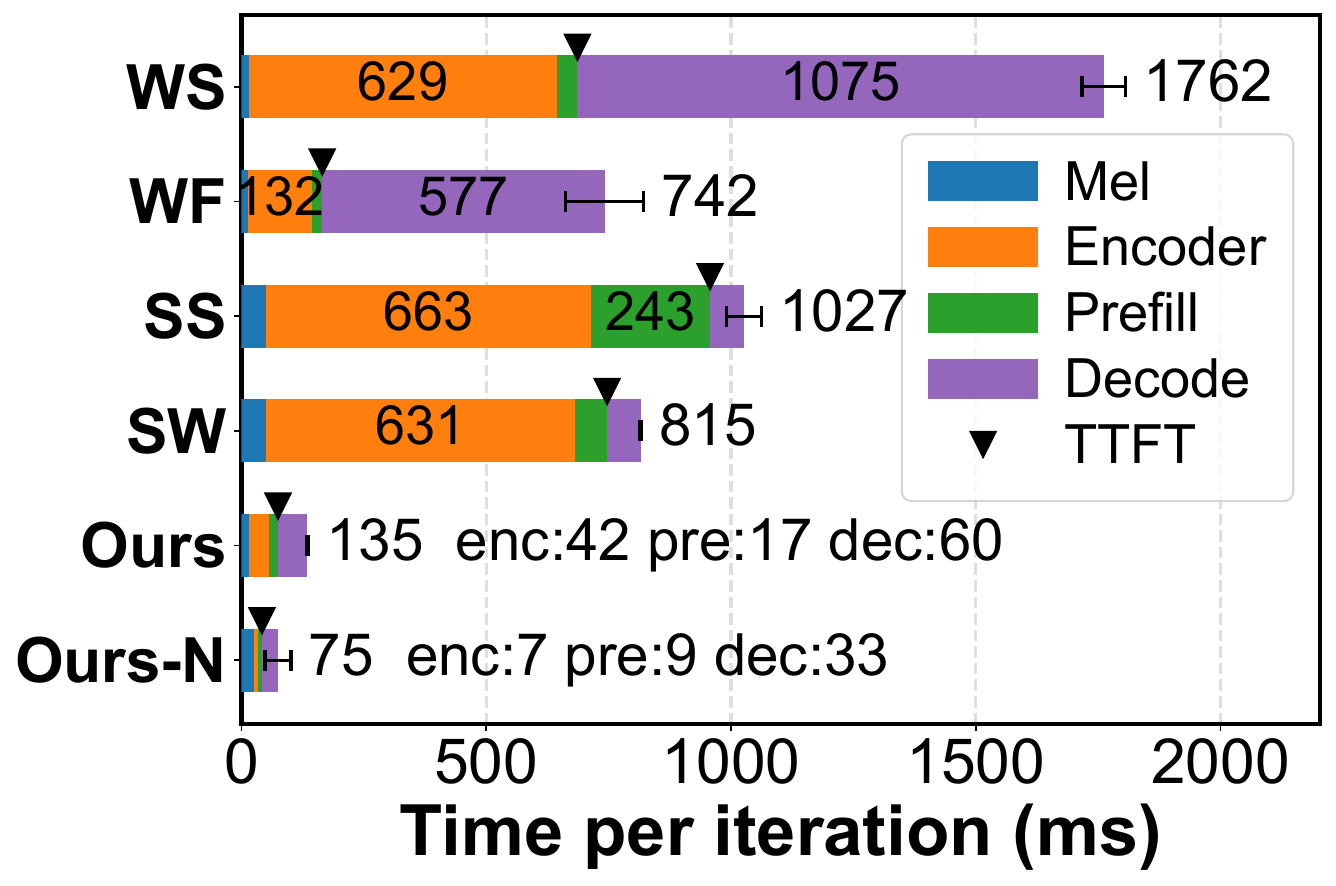}
        \caption{Samsung Galaxy S25}
        \label{fig:latency_breakdown_s25}
    \end{subfigure}
    \begin{subfigure}[t]{0.32\linewidth}
        \centering
        \includegraphics[width=\linewidth]{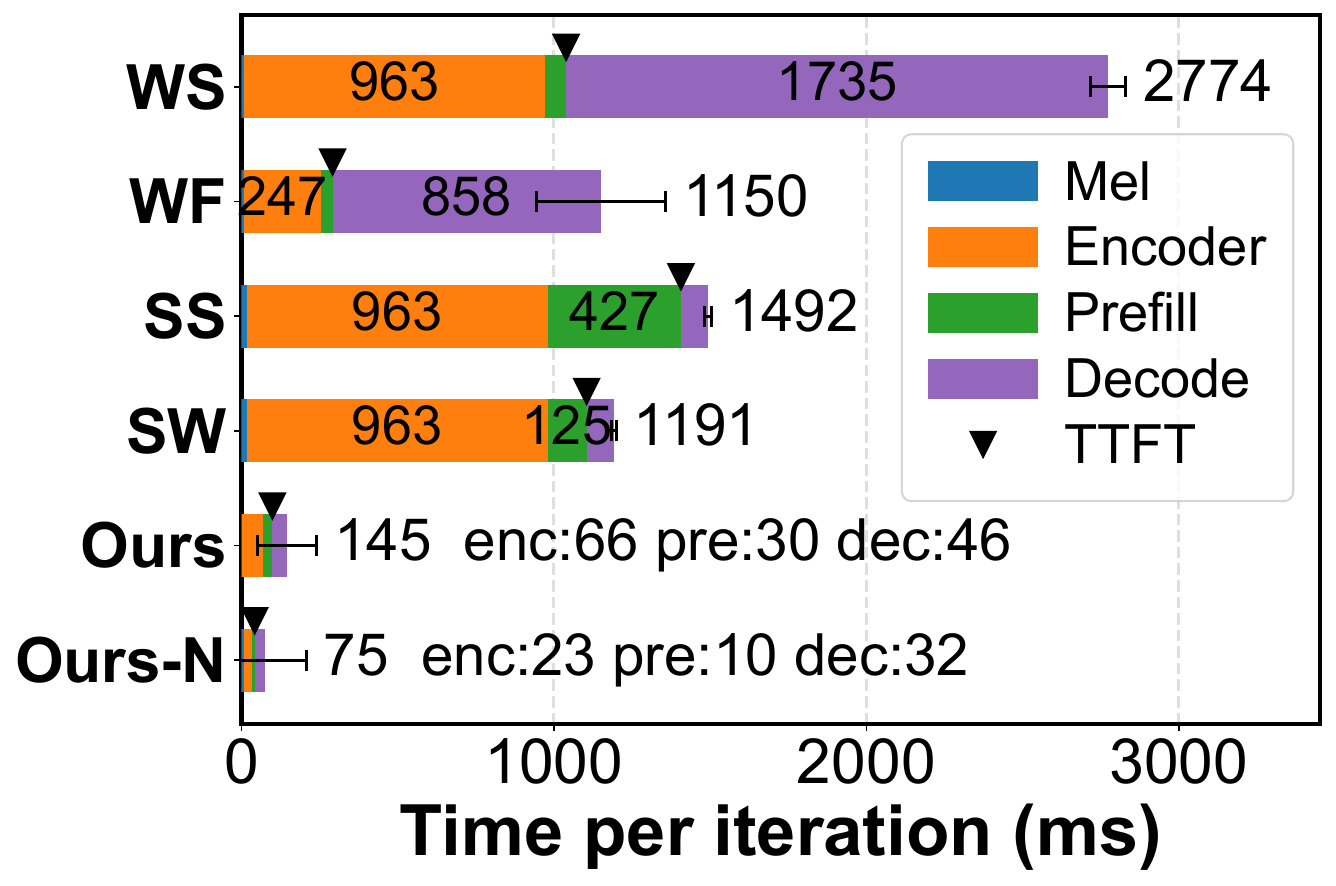}
        \caption{Snapdragon X Plus}
        \label{fig:latency_breakdown_xplus}
    \end{subfigure}
    \caption{Latency breakdown. \sysname delivers the best latency performance by avoiding redundant computation through hallucination detection during decoding (\S~\ref{sec:hallucination-detection}).}
    \label{fig:latency-breakdown}
\end{figure}

\subsection{Latency and Inference Efficiency}
\textbf{Low-latency transcription.}
As shown in Figure~\ref{fig:latency-vs-WER}, \sysname achieves the lowest per-word latency with comparable transcription accuracy, with 1.71--4.62$\times$ and 1.69--4.84$\times$ lower per-word latency than the baselines on S25 and X Plus, respectively.
This gain comes from eliminating redundant computation in padded and repeated Whisper inputs, which reduces inference time, and from using AlignAtt~\citep{papi2023alignatt} to emit tokens from a single inference.

\textbf{Reduced inference cost.}
Figure~\ref{fig:latency-breakdown} shows that \sysname runs Whisper inference 5.52--13.09$\times$ and 7.92--19.10$\times$ faster than the baselines on S25 and X Plus.
\sysname achieves this by avoiding unnecessary padding and processing only newly arrived audio with minimal carryover, resulting in shorter Whisper inputs and lower encoder latency.
\sysname also reduces prefill latency by using only the last valid word as the decoder prefix.

\textbf{Fast transcript updates.}
By avoiding redundant computation during each inference, \sysname also
reduces the time required to produce the first visible transcript update. On
S25, \sysname reduces TTFT by 2.2--12.8$\times$ compared with the baselines,
while Ours-N further improves this to 3.9--22.8$\times$. On X Plus, \sysname
reduces TTFT by 2.9--14.2$\times$, and Ours-N achieves a 6.9--33.2$\times$
reduction. This improvement comes from jointly reducing computation cost, hypothesis recomputation, and unnecessary prefill computation.

\subsection{Energy and Decoding Efficiency}
We measure power on the S25 using a Monsoon High Voltage Power Monitor~\citep{monsoon_hvpm} and on the Snapdragon X Plus using Qualcomm Profiler~\citep{qualcomm_profiler}, reporting power after subtracting idle power. (Detailed in Appendix~\ref{app:power-measurement})

\begin{figure}[t]
    \centering
    \setlength{\abovecaptionskip}{2pt}
    \setlength{\belowcaptionskip}{0pt}

    \begin{minipage}[t]{0.66\linewidth}
        \vspace{0pt}
        \centering
        \captionsetup{type=table}
        \caption{Energy and decoding efficiency. Power and energy metrics subtract base power; throughput is active-inference tokens/s.}
        \label{tab:power-energy}

        \scriptsize
        \setlength{\tabcolsep}{1.0pt}
        \renewcommand{\arraystretch}{0.72}
        \resizebox{0.9\linewidth}{!}{
        \begin{tabular}{llcccc}
            \toprule
            \textbf{Dev.} & \textbf{Sys.} &
            \textbf{P (W,$\downarrow$)} &
            \textbf{E/inf (J,$\downarrow$)} &
            \textbf{E/tok (J,$\downarrow$)} &
            \textbf{Tput (tok/s,$\uparrow$)} \\
            \midrule
            S25 & WS     & 4.05$\pm$0.10 & 13.95$\pm$1.23 & 0.26$\pm$0.01 & 29.20$\pm$1.10 \\
                & WF     & 3.28$\pm$0.09 & 8.28$\pm$0.27  & 0.15$\pm$0.01 & 83.80$\pm$2.90 \\
                & SS     & 3.04$\pm$0.28 & 6.33$\pm$0.62  & 0.61$\pm$0.04 & 10.10$\pm$0.80 \\
                & SW     & 2.94$\pm$0.01 & 6.13$\pm$0.03  & 0.56$\pm$0.01 & 13.10$\pm$0.20 \\
                & Ours   & 0.56$\pm$0.04 & 1.16$\pm$0.09  & 0.13$\pm$0.01 & 64.40$\pm$0.20 \\
                & Ours-N & \textbf{0.46$\pm$0.02} & \textbf{0.92$\pm$0.05} & \textbf{0.10$\pm$0.01} & \textbf{112.90$\pm$0.60} \\
            \midrule
            X Plus & WS     & 4.73$\pm$1.50 & 16.90$\pm$5.37 & 0.30$\pm$0.10 & 18.02$\pm$0.18 \\
                   & WF     & 4.82$\pm$1.37 & 11.21$\pm$3.18 & 0.21$\pm$0.06 & 50.75$\pm$1.26 \\
                   & SS     & 3.27$\pm$1.43 & 6.77$\pm$2.95  & 0.64$\pm$0.28 & 7.25$\pm$0.18 \\
                   & SW     & 2.82$\pm$1.66 & 5.82$\pm$3.42  & 0.52$\pm$0.31 & 9.41$\pm$0.13 \\
                   & Ours   & 1.42$\pm$1.86 & 2.92$\pm$3.81  & 0.33$\pm$0.43 & 64.60$\pm$0.01 \\
                   & Ours-N & \textbf{0.81$\pm$1.75} & \textbf{1.65$\pm$3.57} & \textbf{0.19$\pm$0.41} & \textbf{144.98$\pm$1.09} \\
            \bottomrule
        \end{tabular}
        }
    \end{minipage}
    \hfill
    \begin{minipage}[t]{0.31\linewidth}
        \vspace{15pt}
        \centering
        \includegraphics[
            width=\linewidth,
            height=0.2\textheight,
            keepaspectratio
        ]{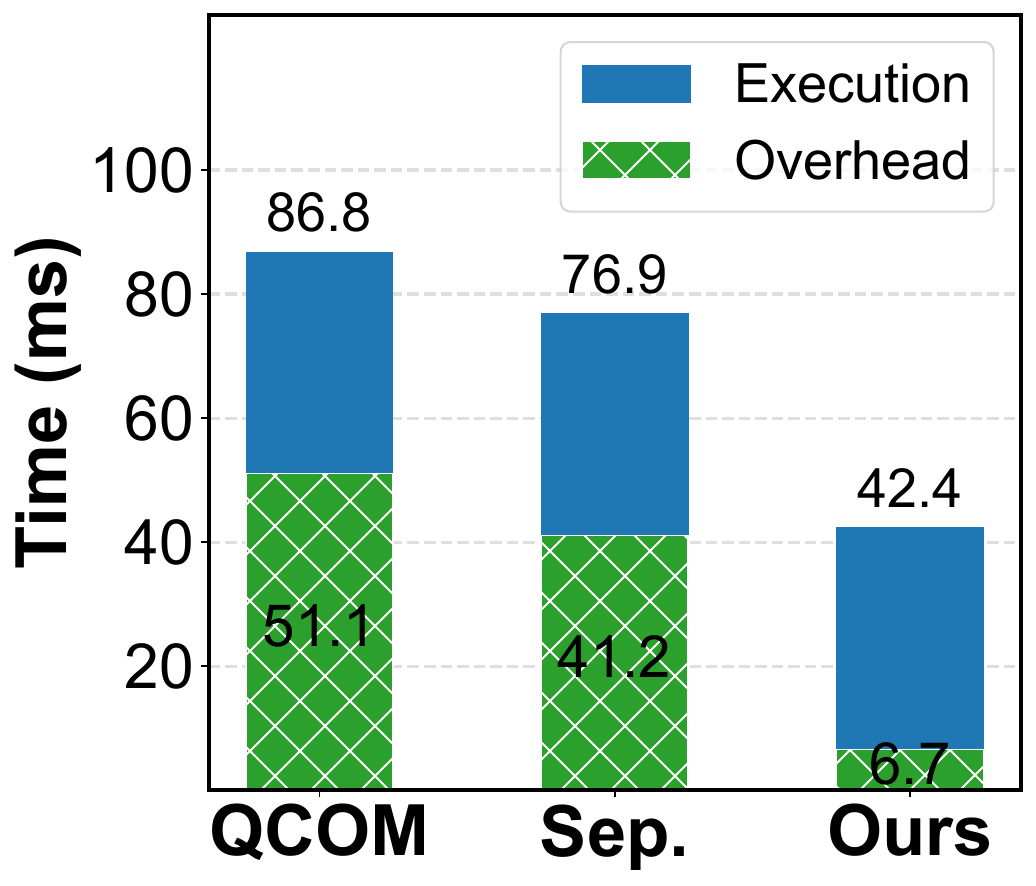}
        \caption{Decoding overhead analysis on S25.}
        \label{fig:decoding-breakdown}
    \end{minipage}
\end{figure}

\textbf{Energy efficiency.}
Table~\ref{tab:power-energy} shows that \sysname achieves the best energy efficiency on both devices.
On S25, Ours-N reduces average power to 0.46 W and energy per inference to 0.92 J, achieving 6.3--8.8$\times$ lower power than the baselines.
On X Plus, Ours-N also achieves the lowest power, energy per inference, and energy per token.
This gain comes from eliminating redundant Whisper computation at the system level, while NPU execution provides additional energy savings. Additional results are provided in Appendix~\ref{app:power-measurement}.

\textbf{Token-metric analysis.}
The energy-per-token and throughput metrics should be interpreted with each decoding policy in mind.
WS and WF sometimes show competitive energy per token or throughput, not because they are more efficient, but because Local Agreement~\citep{liu2020low} regenerates many hypothesis tokens within each inference.
Moreover, WS and WF failed to process the entire audio (see truncated power traces in Figure~\ref{fig:power}) because long Whisper inputs and repeated hypothesis recomputation increase per-iteration latency. Slow iterations delay longest-prefix confirmation in Local Agreement, causing the audio buffer to grow beyond what the systems can process in real time.
By contrast, SS, SW, and \sysname all use AlignAtt decoding~\citep{papi2023alignatt}, making their token-generation metrics more directly comparable. \sysname nevertheless achieves the best energy-per-token and throughput overall.

\textbf{Decoding efficiency.}
Figure~\ref{fig:decoding-breakdown} compares three NPU decoding strategies on S25.
Qualcomm's fixed full-length graph incurs the highest latency due to excessive KV-cache computation.
Using separate graphs removes this redundant KV-cache work but still suffers from per-token graph dispatch overhead.
\sysname reduces both costs by computing only the required KV-cache attention and reducing graph dispatch frequency through controlled unrolling. (Detailed in Appendix~\ref{app:controlled-unrolling-additional-results} for additional results.)

%% file: 6-conclusion.tex
\section{Conclusion and Limitations}

We propose \sysname, an on-device live transcription system that makes Whisper efficient on mobile NPUs by eliminating redundant computation on both audio processing and autoregressive decoding.
\sysname removes the need for long padded Whisper inputs by detecting hallucinated tokens online from cross-attention dynamics, allowing each inference round to process only newly arrived audio with minimal carry-over context.
For NPU execution, \sysname introduces controlled unrolling, which reduces redundant KV-cache attention computation and graph dispatch overhead while enabling chunk-level token emission and decoding control. \sysname achieves the best per-word latency, inference speed, and energy efficiency among the evaluated live transcription systems.

\textbf{Limitations.} This paper focuses on system-level efficiency rather than modifying the Whisper architecture itself. Prior work \citep{krichli2025carelesswhisper} converts Whisper's encoder into a causal encoder and applies low-rank adaptation \citep{hu2022lora}, but the resulting model is limited to the dataset used for adaptation. We focus on building and evaluating a full end-to-end on-device system with the Whisper base model under mobile memory and compute constraints. However, our hallucination detection method is not limited to the base model, as it relies on temporal alignment in Whisper's decoder cross-attention. Appendix~\ref{app:hallucination-generalizability} demonstrates its generalizability across different Whisper model sizes. We leave the design of efficient on-device systems using large speech models such as Canary \citep{sekoyan2025canary} and SeamlessM4T \citep{barrault2023seamlessm4t} for future work.

%% file: 7-appendix.tex
\section{Hallucination Detection Analysis}
\label{app:hallucination-detection-analysis}
\subsection{Filtered Attention-Difference Signal}
\label{app:filtered-attention-difference}

\textbf{Filtered attention.} For each generated token $i$, we compare its final-layer cross-attention with that of the most recent preceding content token $p(i)$.
The raw attention-difference signal is defined as
\begin{equation}
\delta_i(f) = \tilde{a}_i(f) - \tilde{a}_{p(i)}(f),
\label{eq:attention-delta}
\end{equation}
where $f$ denotes the audio frame index.
To reduce spurious spikes near the beginning and end of the audio as shown in Figure~\ref{fig:token-time-alignment}, we apply a median filter followed by a moving-average filter:
\begin{equation}
\hat{\delta}_i(f)
=
\mathrm{MA}_{w_a}
\left(
    \mathrm{Med}_{w_m}(\delta_i)
\right)(f),
\label{eq:filtered-attention-delta}
\end{equation}
where $\mathrm{Med}_{w_m}$ denotes a median filter with window size $w_m$, and $\mathrm{MA}_{w_a}$ denotes a moving-average filter with window size $w_a$.
We use the filtered signal $\hat{\delta}_i(f)$ to locate the positive and negative peaks used for hallucination detection.

\subsection{Component Ablation Study}\label{app:hallucination-ablation}
\textbf{Effect of token-validation components.} Table~\ref{tab:token-validation-ablation} analyzes the effect of each component in our token validation method using 250 samples from the LibriSpeech dataset~\citep{panayotov2015librispeech}.
Without padding, Whisper produces a high hallucination rate.
Backward-shift detection substantially reduces hallucinations, while smoothing and median filtering reduce undesired backward-shift detections.
Skipping punctuation and subword tokens further improves robustness by avoiding unreliable temporal anchors.
The full configuration eliminates hallucinations in this study and achieves the lowest undesired backward-shift detection rate. In \sysname, even if a rare undesired backward shift stops decoding early, the remaining audio is retained as carryover and processed together with newly arrived audio in the next inference round.

\begin{table}[h]
    \centering
    \caption{Ablation study of token validation components. Hall. denotes hallucination rate. Bwd. Shift denotes backward attention shift detection rate. Undesired Bwd. denotes undesired backward-shift detections.}
    \label{tab:token-validation-ablation}
    \footnotesize
    \setlength{\tabcolsep}{3pt}
    \renewcommand{\arraystretch}{0.9}
    \resizebox{0.98\linewidth}{!}{
    \begin{tabular}{lcrrr}
        \toprule
        \textbf{Method variant} & \textbf{Padding} & \textbf{Hall. (\%)} & \textbf{Bwd. Shift (\%)} & \textbf{Undesired Bwd. (\%)} \\
        \midrule
        30s padded baseline                   & 30s  & 0.0  & --   & --   \\
        No-padding baseline                   & None & 34.4 & --   & --   \\
        + backward-shift detection            & None & 4.4  & 70.8 & 75.7 \\
        + smoothing                           & None & 2.8  & 65.2 & 82.2 \\
        + median filtering                    & None & 1.2  & 51.6 & 67.4 \\
        + skip punctuation tokens             & None & 0.0  & 39.6 & 19.2 \\
        + skip punctuation and subword tokens & None & 0.0  & 36.4 & 9.9  \\
        \bottomrule
    \end{tabular}
    }
\end{table}

\begin{figure}[b]
    \centering
    \begin{subfigure}[t]{0.48\linewidth}
        \centering
        \includegraphics[width=\linewidth]{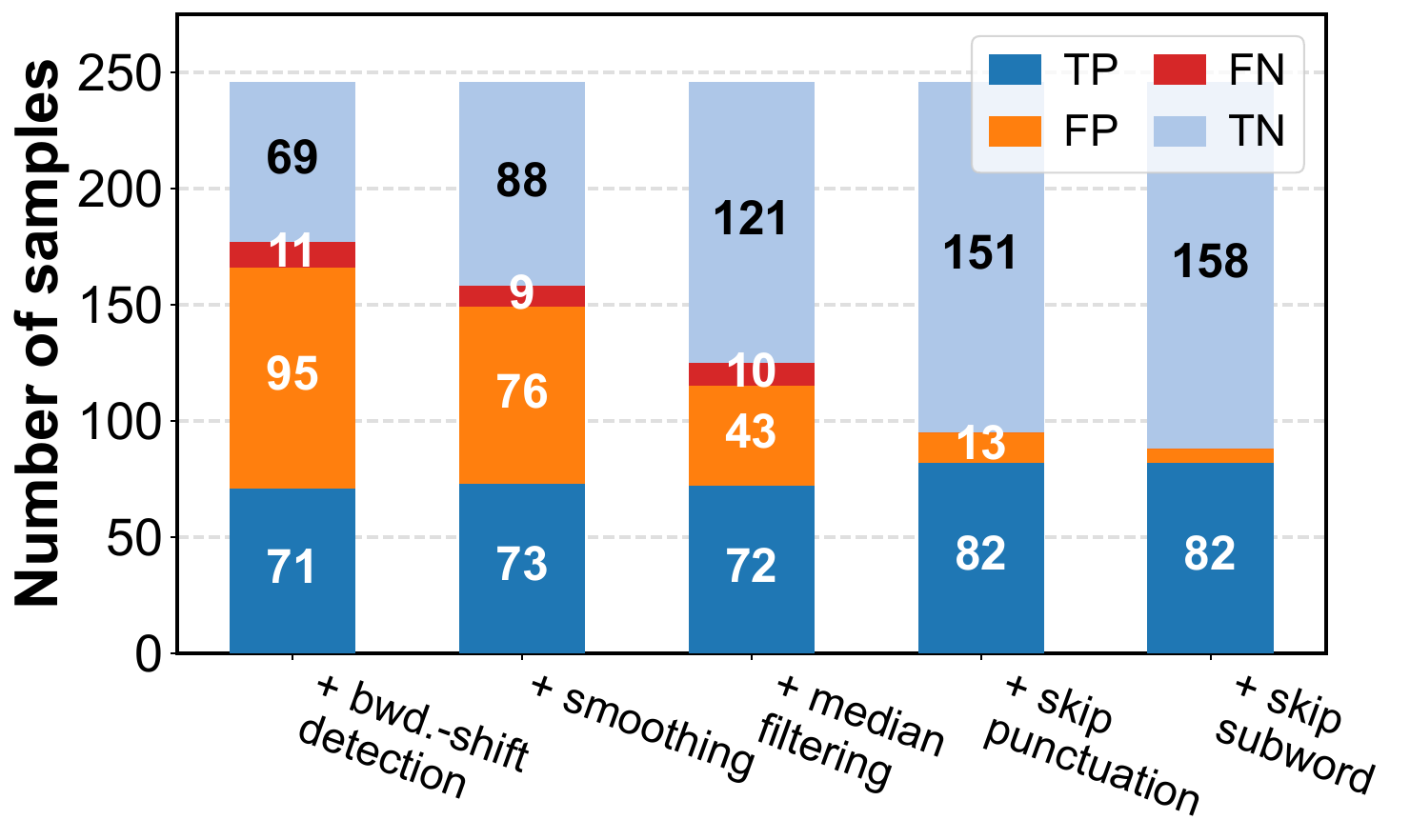}
        \caption{Component-level outcome breakdown}
        \label{fig:ablation-stacked-bar}
    \end{subfigure}
    \hfill
    \begin{subfigure}[t]{0.48\linewidth}
        \centering
        \includegraphics[width=\linewidth]{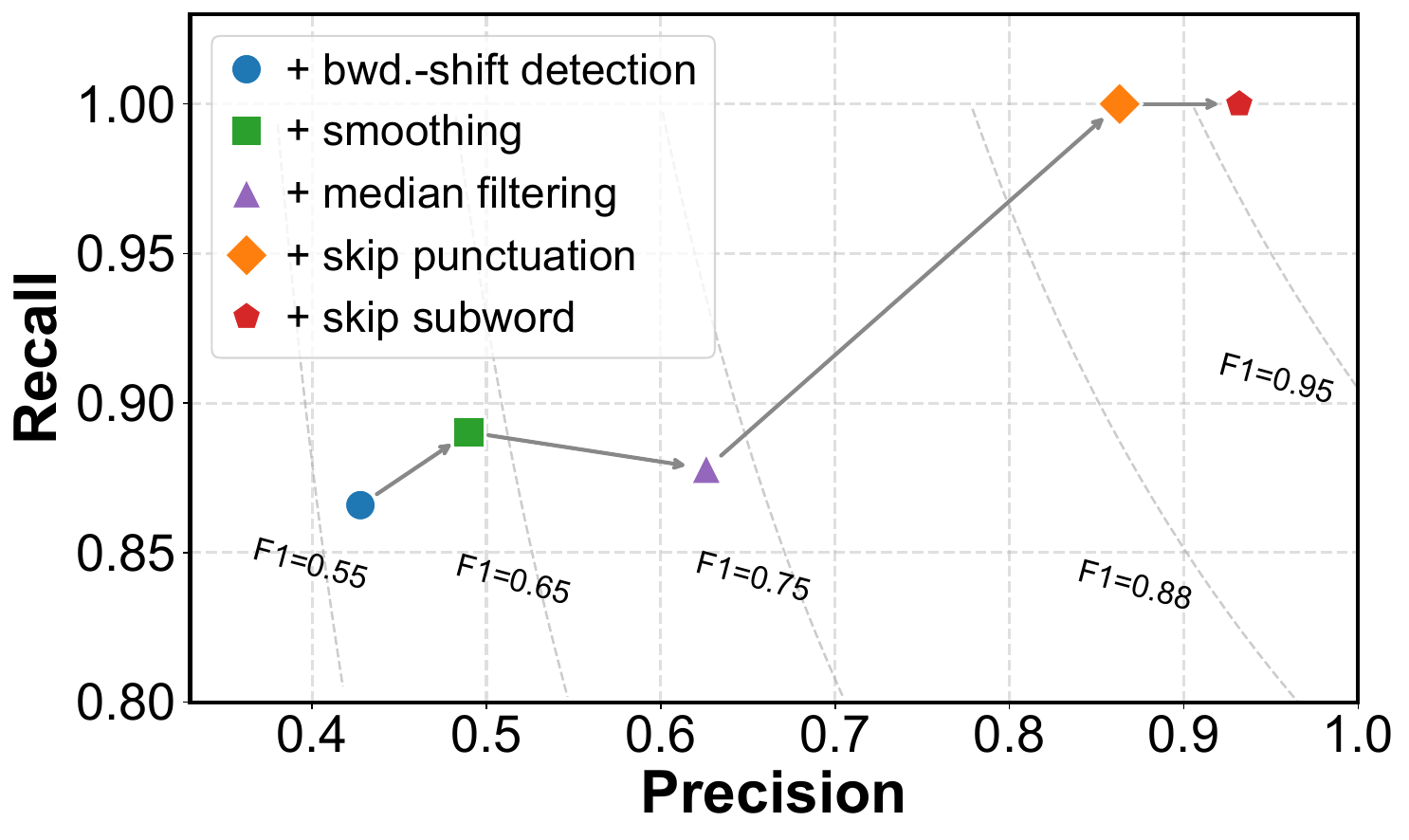}
        \caption{Precision and recall}
        \label{fig:ablation-precision-recall}
    \end{subfigure}
    \caption{Additional analysis of hallucination detection. The stacked bar plot summarizes the outcome distribution for each token-validation variant, while the precision--recall plot characterizes detection quality under different filtering and token-skipping choices.}
    \label{fig:hallucination-ablation-analysis}
\end{figure}

\textbf{Ablation visualization.} Figure~\ref{fig:hallucination-ablation-analysis} provides a complementary view of the same ablation.
The stacked bars (Figure~\ref{fig:hallucination-ablation-analysis}(~\subref{fig:ablation-stacked-bar})) show that each component progressively reduces false or undesired detections while preserving the ability to catch hallucinated tokens.
The precision--recall plot (Figure~\ref{fig:hallucination-ablation-analysis}(~\subref{fig:ablation-precision-recall})) further shows that token filtering improves detection precision while maintaining high recall, indicating that punctuation and subword tokens are unreliable temporal anchors for hallucination detection.

\subsection{Filter Parameter Selection}
\label{app:filter-parameter-selection}

\textbf{Parameter selection rationale.} We choose the moving-average and median-filter window sizes in two steps.
Table~\ref{tab:filter-parameter-sweep} reports representative configurations used for this selection.
First, without median filtering, we vary the moving-average window and find that $w_a=10$ reduces the hallucination rate among the tested smoothing settings.
Second, with $w_a=10$ fixed, we add a median filter to remove spurious attention peaks in the attention-difference signal, which are visible in Figure~\ref{fig:token-time-alignment}.
A median-filter width of $w_m=7$ removes most of these peaks and reduces undesired backward-shift detections while maintaining a low hallucination rate.
We therefore use $(w_a, w_m)=(10,7)$ in our implementation.
Compared with this parameter selection, the larger improvement comes from the backward-shift criterion and proper handling of punctuation and subword tokens, as shown in Table~\ref{tab:token-validation-ablation} and Figure~\ref{fig:hallucination-ablation-analysis}.

\begin{table}[h]
    \centering
    \caption{Filter parameter selection. Hall., Bwd. Shift, and Undesired Bwd. are reported in percentage.}
    \label{tab:filter-parameter-sweep}
    \footnotesize
    \setlength{\tabcolsep}{5pt}
    \renewcommand{\arraystretch}{0.9}
    \resizebox{0.6\linewidth}{!}{
    \begin{tabular}{lrrr}
        \toprule
        \textbf{Config. (MA, Med.)} &
        \textbf{Hall.} &
        \textbf{Bwd. Shift} &
        \textbf{Undesired Bwd.} \\
        \midrule
        (0, 0)  & 4.4 & 70.8 & 75.7 \\
        (5, 0)  & 3.6 & 62.0 & 83.2 \\
        (10, 0) & 2.8 & 65.2 & 82.2 \\
        (15, 0) & 3.2 & 67.6 & 84.0 \\
        \midrule
        (10, 3) & 2.4 & 60.0 & 74.7 \\
        (10, 5) & 1.2 & 59.2 & 73.0 \\
        (10, 7) & 1.2 & 51.6 & 67.4 \\
        \bottomrule
    \end{tabular}
    }
\end{table}

\subsection{Model Selection and Generalizability}
\label{app:hallucination-generalizability}

\textbf{Model selection.} Our end-to-end system evaluation focuses on the Whisper base model. This choice reflects the memory and compute constraints of mobile devices, where the goal is to build a complete real-time on-device transcription system rather than to maximize transcription accuracy with a larger model. We exclude Whisper tiny because its transcription quality is substantially weaker and it is rarely used as the main model in recent Whisper-based streaming systems~\citep{wang2025whisperflow}. We also exclude Whisper large from this analysis in this section because it caused an out-of-memory error even on the Galaxy S25, which is a high-end mobile device.

\textbf{Cross-model evaluation.}
Although our end-to-end system uses Whisper base, the proposed hallucination detection method (\S~\ref{sec:hallucination-detection}) is not specific to the base model. Its generality comes from relying on temporal alignment in decoder cross-attention, which is consistently observed across different Whisper model sizes. To evaluate this property, we reuse the configuration chosen for Whisper base on Whisper small and medium, including the same filtering and token-skipping parameters. For each model size, we compare a no-detection baseline against our method.

\begin{table}[h]
    \centering
    \caption{Generalizability of hallucination detection across Whisper model sizes. Hall. denotes hallucination rate. Bwd. denotes backward attention shift detection rate. Undesired denotes undesired backward-shift detections. All ``Our method'' rows use the same configuration selected for Whisper base.}
    \label{tab:model-size-generalizability}
    \small
    \setlength{\tabcolsep}{4pt}
    \renewcommand{\arraystretch}{0.95}
    \begin{tabular}{llcccc}
        \toprule
        \textbf{Model} &
        \textbf{Config.} &
        \textbf{Layer} &
        \textbf{Hall. (\%)} &
        \textbf{Bwd. (\%)} &
        \textbf{Undesired (\%)} \\
        \midrule
        Small  & No detection & -- & 24.0 & --   & --   \\
               & Our method   & 5  & 2.0  & 5.2  & 7.7  \\
               & Our method   & 7  & 2.0  & 3.2  & 12.5 \\
        \midrule
        Medium & No detection & -- & 36.4 & --   & --   \\
               & Our method   & 8  & 6.8  & 28.8 & 6.9  \\
        \bottomrule
    \end{tabular}
\end{table}

\textbf{Generalizability.} Table~\ref{tab:model-size-generalizability} shows that hallucination detection remains effective beyond Whisper base. On Whisper small, our method reduces the hallucination rate from 24.0\% to 2.0\% for both selected layers. Undesired backward-shift remains modest at 5.2\% and 3.2\%, respectively. On Whisper medium, our method reduces the hallucination rate from 36.4\% to 6.8\%, with 6.9\% undesired backward-shift. These results indicate that temporal alignment in decoder cross-attention provides a general signal for detecting hallucinated tokens across different Whisper model sizes.

\textbf{Deployment feasibility.} Although our hallucination detection method remains effective, Whisper medium is less practical for real-time deployment because its encoder alone takes more than 4~s on the Snapdragon X Plus laptop, which exceeds the short inference interval in live transcription systems. This makes Whisper medium difficult to use as a full real-time on-device system, despite its potential for higher transcription accuracy. We leave the construction of fully optimized end-to-end systems for larger Whisper model sizes to future work, including model-size-specific hyperparameter tuning for hallucination detection.

\section{Additional Results}\label{app:additional-results}

\subsection{Transcription Quality}

\textbf{Word and character errors.} Table~\ref{tab:additional-system-wer-cer} reports additional transcription
quality results. All systems are evaluated
with the Whisper base model, and WER/CER are computed per sample and reported as
mean$\pm$standard deviation. Full WER/CER compare each system output
against the complete reference transcript. For Whisper-Streaming and WhisperFlow, some runs terminate before
processing the full audio, which can inflate full-reference WER/CER due to
incomplete coverage. We therefore additionally report truncated WER/CER for
these cases, comparing each output only against the reached reference prefix.

\begin{center}
    \centering
    \captionof{table}{Additional per-sample full and truncated WER/CER across systems. WS, WF, SS, and SW denote Whisper-Streaming, WhisperFlow, SimulStreaming, and Simul-Whisper, respectively. Values are percentages and reported as mean$\pm$population standard deviation over samples.}
    \label{tab:additional-system-wer-cer}
    \small
    \setlength{\tabcolsep}{5pt}
    \renewcommand{\arraystretch}{0.9}
    \begin{tabular}{llcccc}
        \toprule
        \textbf{Dataset} &
        \textbf{System} &
        \textbf{Full WER} &
        \textbf{Full CER} &
        \textbf{Trunc. WER} &
        \textbf{Trunc. CER} \\
        \midrule
        Meanwhile & WS   & 30.69$\pm$24.23 & 24.89$\pm$25.07 & 11.79$\pm$6.32  & 5.29$\pm$8.62 \\
                  & WF   & 25.13$\pm$22.95 & 18.69$\pm$23.89 & 16.64$\pm$19.56 & 9.58$\pm$20.72 \\
                  & SS   & 12.06$\pm$4.00  & 5.40$\pm$1.89   & \multicolumn{1}{c}{--} & \multicolumn{1}{c}{--} \\
                  & SW   & 12.93$\pm$4.49  & 5.85$\pm$2.20   & \multicolumn{1}{c}{--} & \multicolumn{1}{c}{--} \\
                  & Ours & 16.36$\pm$11.21 & 8.44$\pm$6.29   & \multicolumn{1}{c}{--} & \multicolumn{1}{c}{--} \\
        \midrule
        TED-LIUM 3 & WS   & 91.15$\pm$5.16  & 90.09$\pm$5.73  & 11.04$\pm$3.56  & 2.81$\pm$1.08 \\
                    & WF   & 90.25$\pm$18.95 & 89.63$\pm$19.70 & 32.73$\pm$41.22 & 29.04$\pm$43.47 \\
                    & SS   & 11.26$\pm$3.34  & 3.36$\pm$1.11   & \multicolumn{1}{c}{--} & \multicolumn{1}{c}{--} \\
                    & SW   & 11.40$\pm$2.90  & 3.53$\pm$1.12   & \multicolumn{1}{c}{--} & \multicolumn{1}{c}{--} \\
                    & Ours & 14.18$\pm$2.94  & 5.46$\pm$1.26   & \multicolumn{1}{c}{--} & \multicolumn{1}{c}{--} \\
        \bottomrule
    \end{tabular}
\end{center}

\subsection{Power Measurement}\label{app:power-measurement}

\textbf{Measurement setup.}
We measured full-system battery-rail power on the Galaxy S25 using a Monsoon High Voltage Power Monitor~\citep{monsoon_hvpm}, configured as an external supply for the phone battery rail~\citep{laskaridis2024melting}.
The S25 traces used for Table~\ref{tab:power-energy} were captured with $V_{\mathrm{out}}=4.42$~V.
The Monsoon capture process starts before each live-transcription run and samples the main rail current and voltage at $f_s=5000$~Hz.
The workload script writes start and end marker files immediately before and after the transcription command; all S25 power statistics are computed over the Monsoon samples whose timestamps fall inside this marker-delimited command window.
This measures the complete phone system during the workload, not an isolated GPU or NPU rail.

\textbf{Power and energy computation.} For each command-window sample $i$, we compute instantaneous power as
\begin{equation}
p_i = I_i V_i,
\quad
\bar{P}_{\mathrm{raw}} = \frac{1}{M}\sum_{i=1}^{M} p_i,
\quad
E_{\mathrm{raw}} = \frac{1}{f_s}\sum_{i=1}^{M} p_i,
\label{eq:s25-power-integration}
\end{equation}
where $I_i$ is in mA, $V_i$ is in V, $p_i$ is in mW, and $M$ is the number of samples in the command window.
Thus, S25 energy is obtained by discrete integration over Monsoon samples rather than by manually multiplying a reported average by wall-clock time.
We separately measure an idle baseline with the same Monsoon setup and subtract its command-window mean power, $P_{\mathrm{idle}}=362.2$~mW, from each S25 workload:
\begin{equation}
P_{\mathrm{work}} = \bar{P}_{\mathrm{raw}} - P_{\mathrm{idle}},
\quad
E_{\mathrm{work}} = E_{\mathrm{raw}} - P_{\mathrm{idle}}\frac{M}{f_s}.
\label{eq:s25-idle-subtraction}
\end{equation}

\textbf{Reported metrics.} Table~\ref{tab:power-energy} reports $P_{\mathrm{work}}$ and energy metrics derived from $E_{\mathrm{work}}$, converted to W and J.
Energy per inference is $E_{\mathrm{work}}/N$, where $N$ is the number of inference iterations in the run logs.
Energy per token divides $E_{\mathrm{work}}$ by the total number of generated tokens, and throughput is computed as generated tokens per active inference time, where active time is the sum of mel, encoder, decoder-prefill, and decoder execution time.

\begin{figure}[h]
    \centering
    \setlength{\abovecaptionskip}{2pt}
    \setlength{\belowcaptionskip}{2pt}

    \begin{subfigure}[t]{0.32\linewidth}
        \centering
        \vspace{0pt}
        \includegraphics[width=\linewidth]{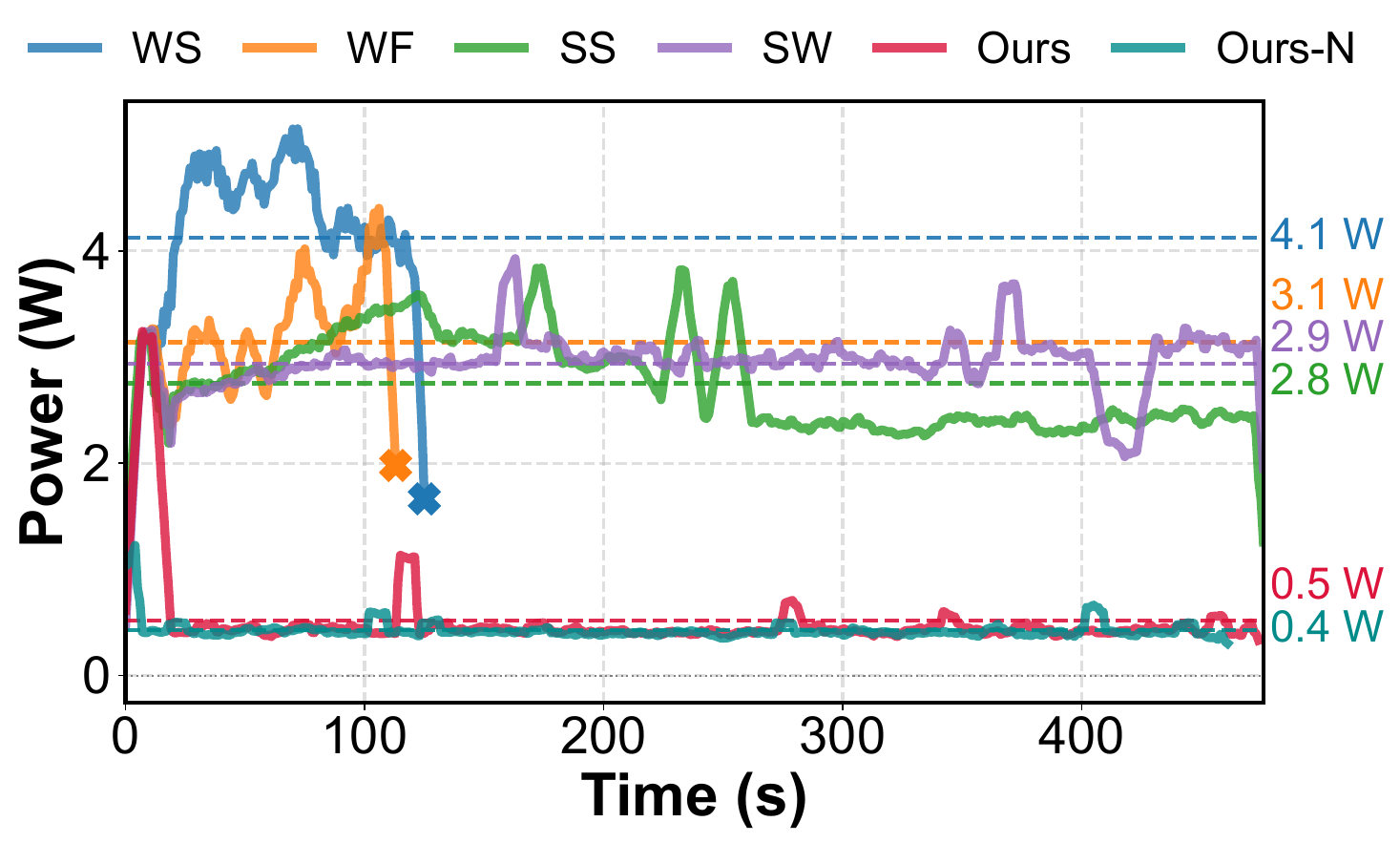}
        \caption{EricMead on S25}
        \label{fig:power-s25-eric}
    \end{subfigure}
    \hfill
    \begin{subfigure}[t]{0.32\linewidth}
        \centering
        \vspace{0pt}
        \includegraphics[width=\linewidth]{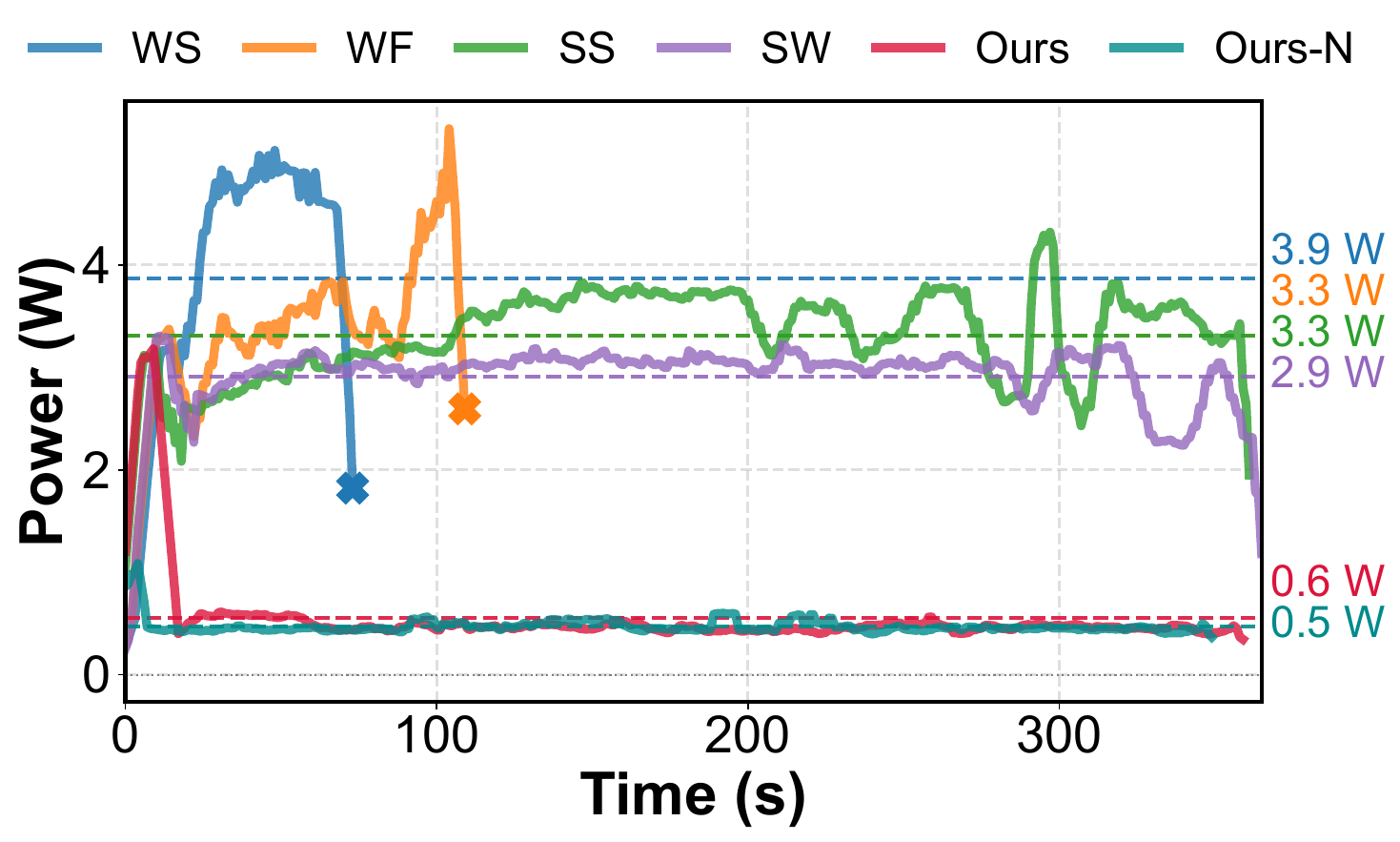}
        \caption{GaryFlake on S25}
        \label{fig:power-s25-gary}
    \end{subfigure}
    \hfill
    \begin{subfigure}[t]{0.32\linewidth}
        \centering
        \vspace{0pt}
        \includegraphics[width=\linewidth]{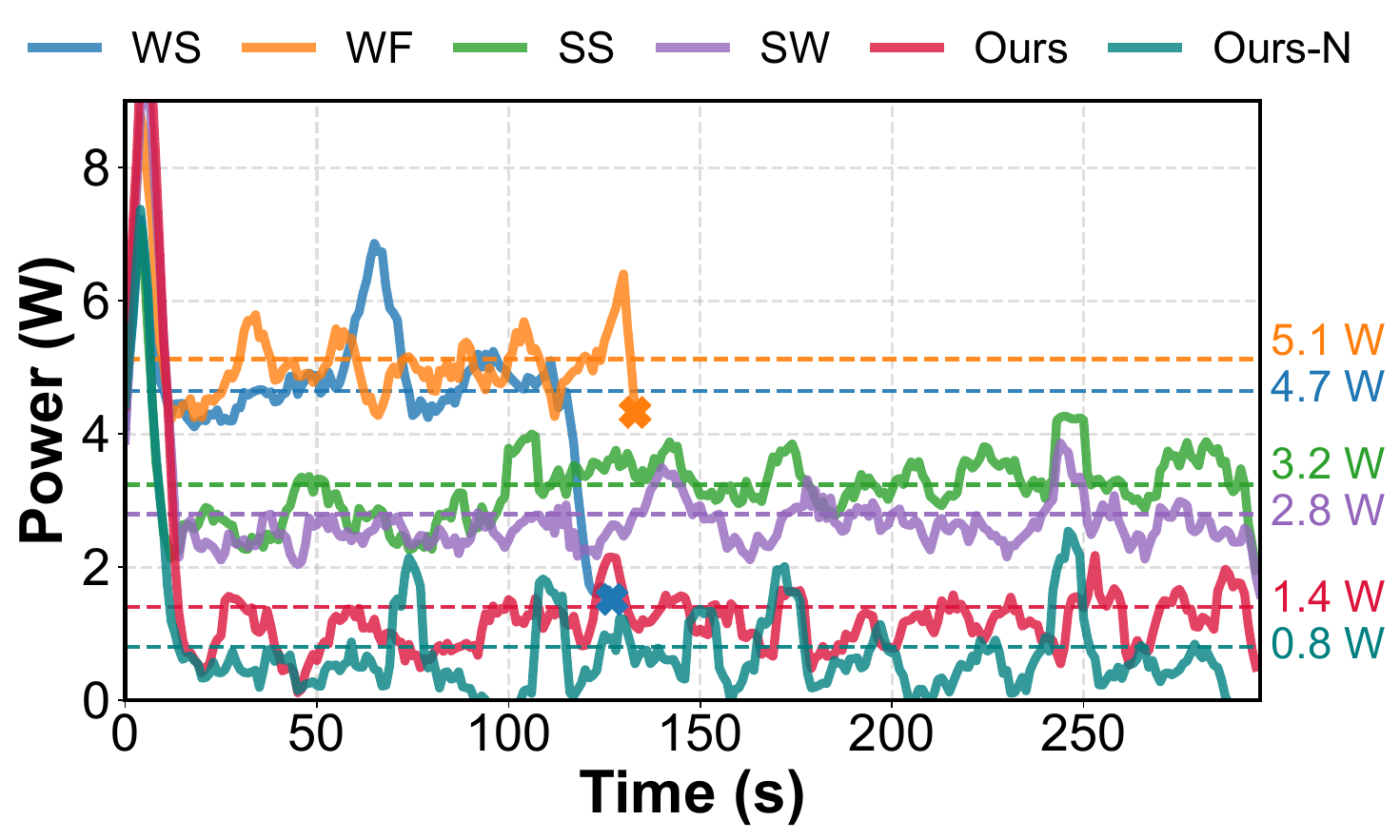}
        \caption{GaryFlake on X Plus}
        \label{fig:power-xplus-gary}
    \end{subfigure}

    \caption{Power consumption. \sysname achieves the lowest power consumption by reducing redundant computation on both GPU and NPU. The \(\times\) markers indicate when each system terminates.}
    \label{fig:power}
\end{figure}

\textbf{Snapdragon X Plus measurement.} For the Snapdragon X Plus laptop, we measured power using Qualcomm Profiler~\citep{qualcomm_profiler} and applied the same base-power subtraction principle to the profiler traces.
Because the S25 and X Plus use different power measurement interfaces, their traces have different noise characteristics.
We therefore use the traces in Figure~\ref{fig:power} to show qualitative results and report the averaged values in Table~\ref{tab:power-energy} for quantitative comparison.
The numerical values in Table~\ref{tab:power-energy} are computed from the raw command-window summaries.

\textbf{Power traces.} 
Figure~\ref{fig:power} shows the power traces used for Table~\ref{tab:power-energy}.
\sysname consistently maintains much lower power than the baselines because it avoids long padded inputs and repeated hypothesis recomputation.
All systems show an initial power peak when transcription starts, reflecting the startup cost of launching the live transcription pipeline.
The WS and WF traces on both devices are truncated because these systems terminated before processing the entire audio. This behavior is caused by high per-iteration latency from long Whisper inputs and hypothesis recomputation by Local Agreement.

\subsection{Controlled Unrolling on NPUs}\label{app:controlled-unrolling-additional-results}

\begin{figure}[h]
    \centering

    \begin{minipage}[t]{0.24\linewidth}
        \centering
        \includegraphics[
            width=\linewidth,
            height=0.18\textheight,
            keepaspectratio
        ]{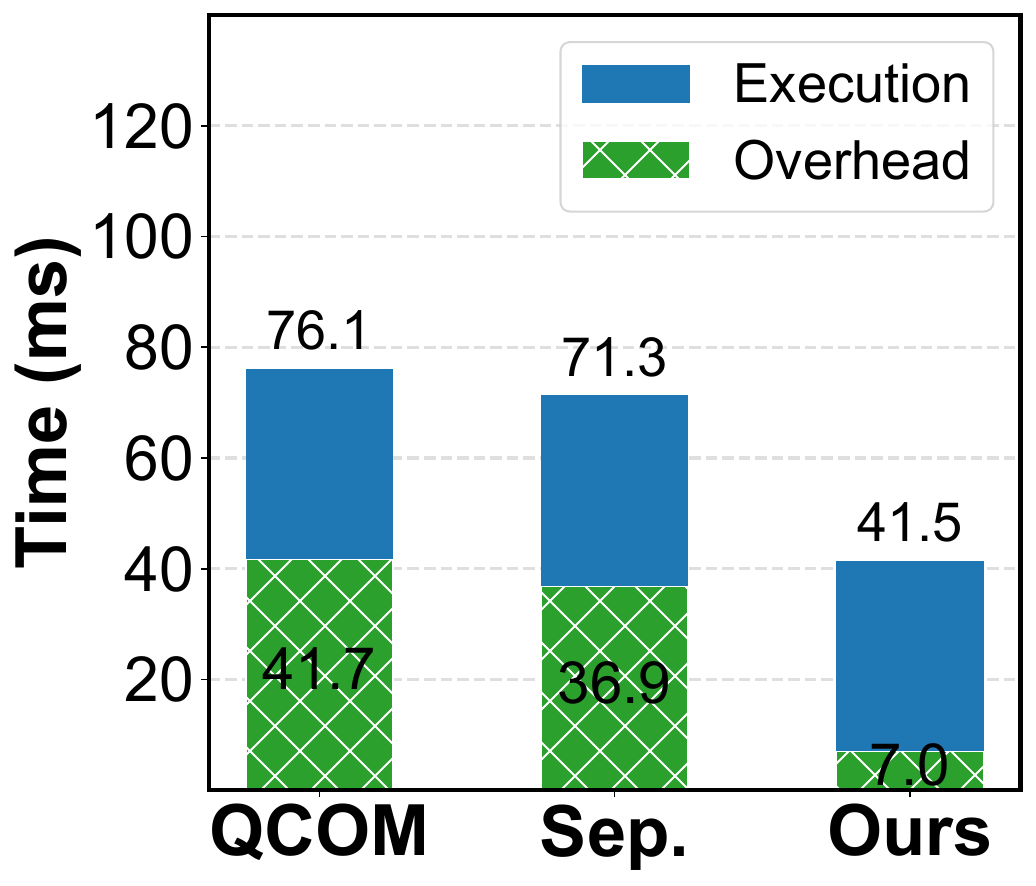}
        \captionof{figure}{Decoding overhead on X Plus.}
        \label{fig:decoding-breakdown-xplus}
    \end{minipage}
    \hfill
    \begin{minipage}[t]{0.74\linewidth}
        \centering

        \begin{minipage}[t]{0.32\linewidth}
            \centering
            \includegraphics[
                width=\linewidth,
                height=0.18\textheight,
                keepaspectratio
            ]{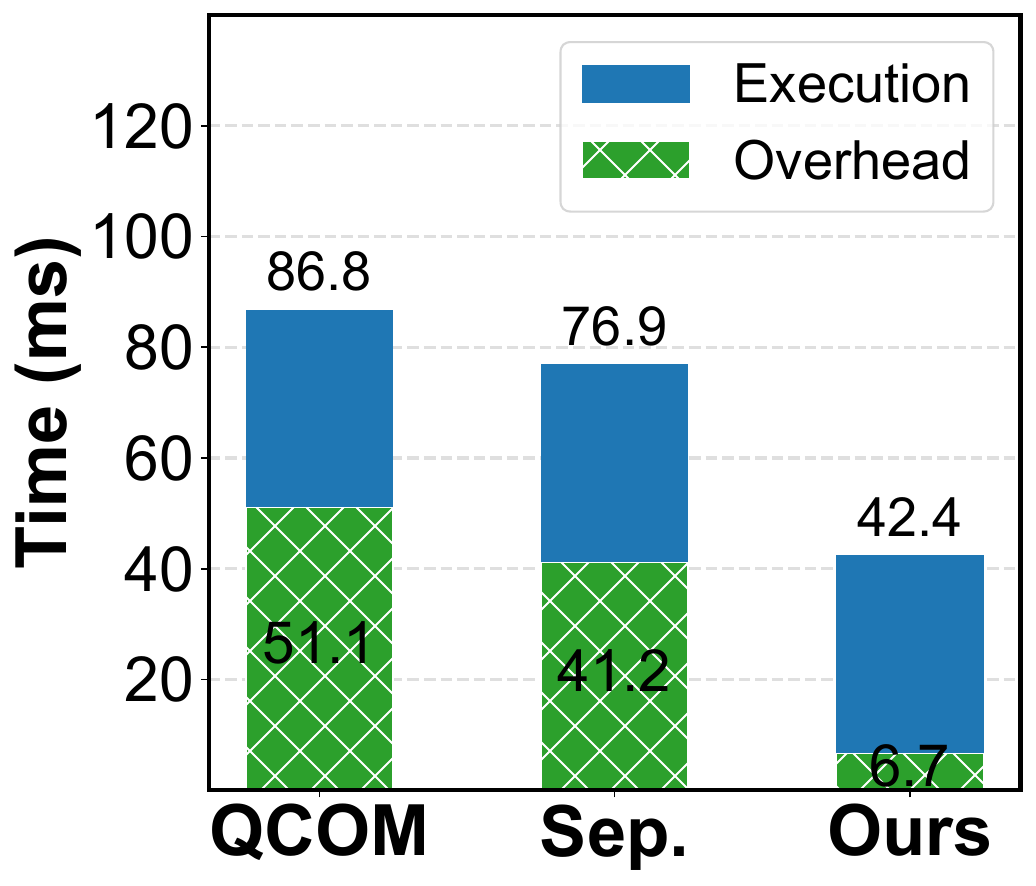}
            \subcaption{16 tokens}
        \end{minipage}
        \hfill
        \begin{minipage}[t]{0.32\linewidth}
            \centering
            \includegraphics[
                width=\linewidth,
                height=0.18\textheight,
                keepaspectratio
            ]{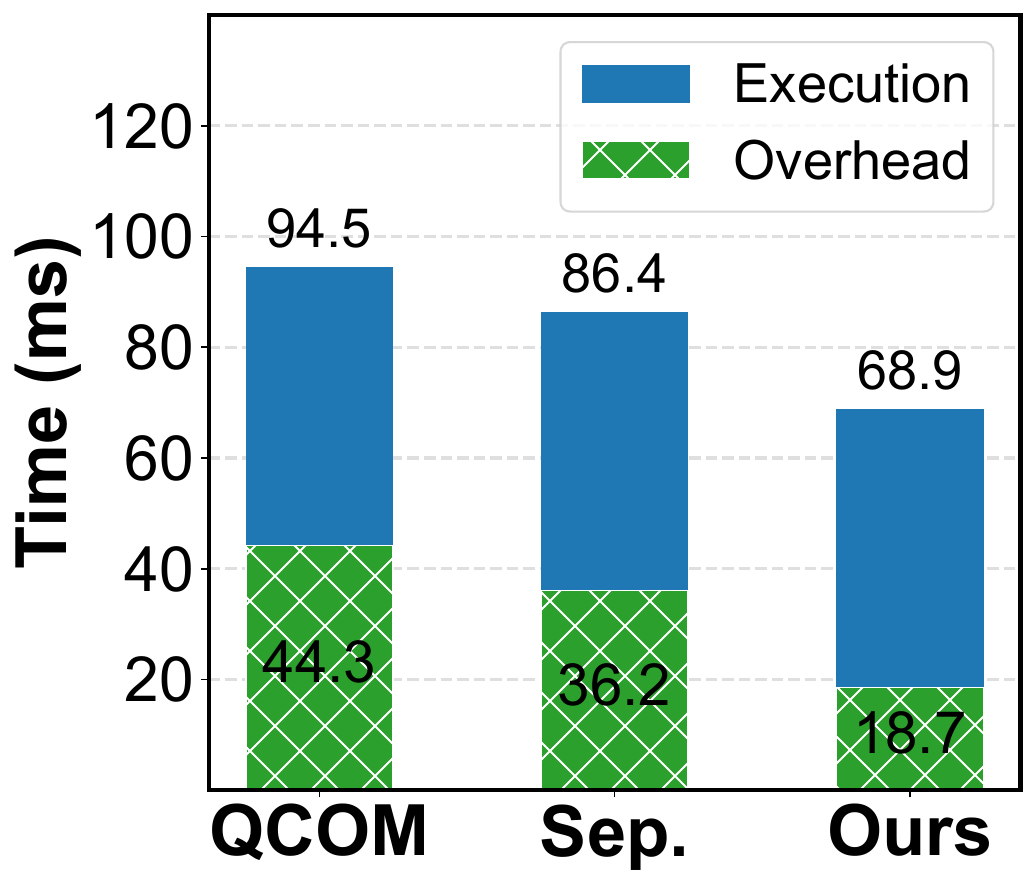}
            \subcaption{23 tokens}
        \end{minipage}
        \hfill
        \begin{minipage}[t]{0.32\linewidth}
            \centering
            \includegraphics[
                width=\linewidth,
                height=0.18\textheight,
                keepaspectratio
            ]{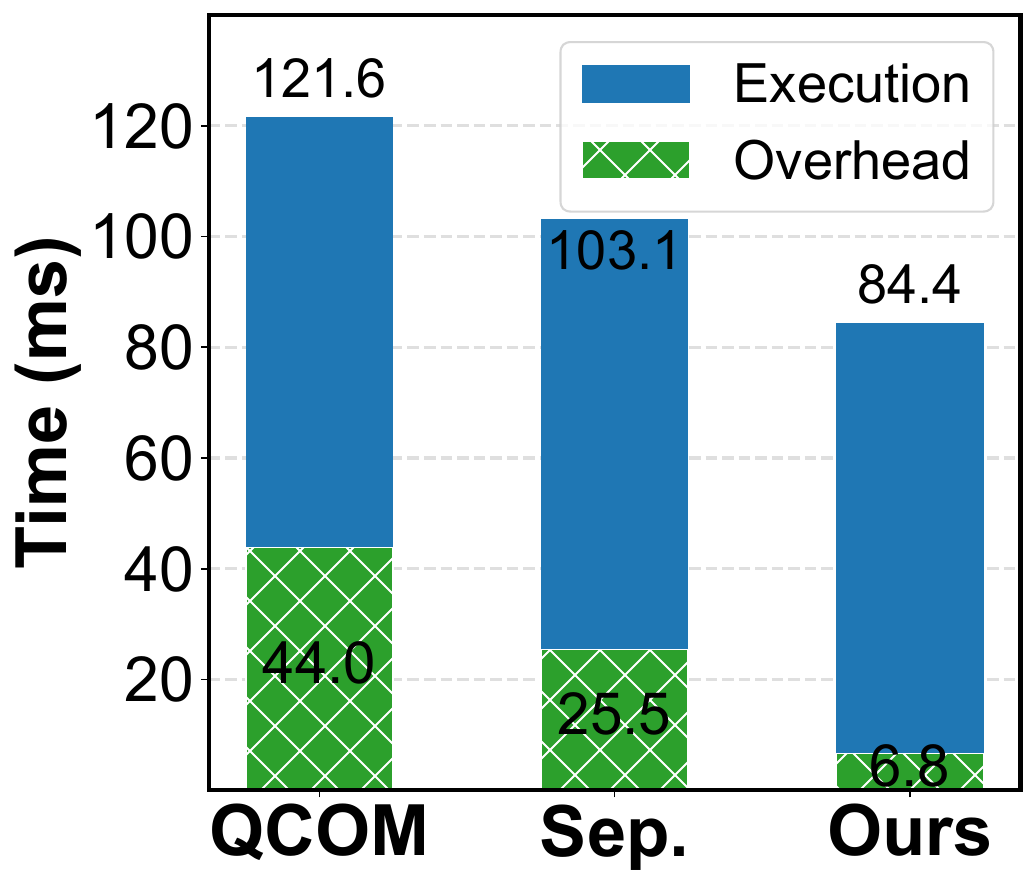}
            \subcaption{30 tokens}
        \end{minipage}
        \captionof{figure}{Decoding overhead analysis on S25 as the number of generated tokens varies.}
        \label{fig:decoding-breakdown-s25-token-counts}
    \end{minipage}

\end{figure}

\textbf{Cross-device consistency.} In addition to the S25 results in Figure~\ref{fig:decoding-breakdown}, we
provide supplementary X Plus results in
Figure~\ref{fig:decoding-breakdown-xplus}. The X Plus results show the same
overall trend as S25: the fixed full-length graph has the highest decoding
overhead, the separate-graph approach reduces redundant KV-cache computation but
still incurs graph-dispatch overhead, and our controlled-unrolling design
achieves the lowest overall decoding overhead.

\textbf{Across decoding lengths.} We also evaluate whether this trend remains consistent as the number of generated tokens changes. Figure~\ref{fig:decoding-breakdown-s25-token-counts} reports the
S25 decoding breakdown for shorter, medium, and longer decoding cases. Across
all token-count settings, the relative behavior of the three strategies remains
unchanged. The fixed full-length graph remains the least efficient, the
separate-graph approach improves over it but retains dispatch overhead, and our
method consistently provides the lowest decoding overhead.

\subsection{Supplementary CPU-only System Comparison}
\textbf{CPU-only setup.} Table~\ref{tab:cpu-stress-test} reports a supplementary CPU-only system
comparison on Galaxy S25 and Snapdragon X Plus. All systems use the same 2~s
inference interval and run without GPU acceleration. This experiment is not
intended as our main comparison setting. Instead, it characterizes how each
system behaves when real-time streaming must be sustained using CPU execution
alone.
\begin{table}[h]
    \centering
    \caption{CPU-only realtime results on GaryFlake and EricMead. All values are reported as averages over the two samples. PWL denotes per-word latency. Input is the encoder input length, Prep. is the mel-spectrogram computation time, Enc. is encoder execution time, and Dec. includes both prefill and autoregressive decoding time.}
    \label{tab:cpu-stress-test}
    \tiny
    \setlength{\tabcolsep}{2pt}
    \renewcommand{\arraystretch}{0.9}
    \resizebox{\linewidth}{!}{
    \begin{tabular}{llccccccc}
        \toprule
        \textbf{Dev.} &
        \textbf{Sys.} &
        \textbf{Full WER (\%)} &
        \textbf{Trunc. WER (\%)} &
        \textbf{PWL (s)} &
        \textbf{Input (s)} &
        \textbf{Prep. (ms)} &
        \textbf{Enc. (ms)} &
        \textbf{Dec. (ms)} \\
        \midrule
        S25 & WS   & 89.71  & 11.34 & 10.09 & 30.00 & 23.4  & 2141.7 & 852.8 \\
            & WF   & 90.51  & 56.71 & 5.40\footnotemark & 12.81 & 24.6  & 854.3  & 430.0 \\
            & SS   & 58.02  & \multicolumn{1}{c}{--} & 95.38 & 30.00 & 168.0 & 2832.6 & 1142.2 \\
            & SW   & 100.00 & \multicolumn{1}{c}{--} & \multicolumn{1}{c}{--} & 30.00 & 49.4  & 2814.1 & 191.6 \\
            & Ours & 13.43  & \multicolumn{1}{c}{--} & 2.00  & 3.67 & 5.6   & 161.6  & 37.1 \\
        \midrule
        X Plus & WS   & 88.19 & 6.17  & 4.91 & 30.00 & 9.6  & 690.7 & 306.0 \\
               & WF   & 85.44 & 17.63 & 3.92 & 13.99 & 8.3  & 230.3 & 154.4 \\
               & SS   & 7.79  & \multicolumn{1}{c}{--} & 3.31 & 30.00 & 19.4 & 698.0 & 336.1 \\
               & SW   & 7.69  & \multicolumn{1}{c}{--} & 3.08 & 30.00 & 17.4 & 640.7 & 99.0 \\
               & Ours & 9.62  & \multicolumn{1}{c}{--} & 1.99 & 3.76 & 3.2  & 52.3  & 18.3 \\
        \bottomrule
    \end{tabular}
    }
\end{table}
\footnotetext{WF per-word latency is computed from GaryFlake only because the EricMead run did not progress long enough for word-latency matching.}

\textbf{Mobile CPU bottlenecks.} The CPU-only results highlight why removing unnecessary computation is important
for on-device streaming inference. On the S25 CPU, systems that repeatedly use a
fixed 30s encoder input incur substantial encoder overhead: WS, SS, and SW spend
2141.7 ms, 2832.6 ms, and 2814.1 ms, respectively, on encoder execution alone.
These costs already exceed the 2 s inference interval before accounting for
preprocessing and decoding. SS further spends 1142.2 ms in the decoder, which is
the largest decoder cost among the S25 runs, mainly because it performs the most
prefill-token computation. As a result, SS completes but falls far behind the
real-time latency performance, while SW fails to produce a usable transcript.

\textbf{Effect of removing redundant computation.} \sysname avoids much of this overhead by using online hallucination
detection to remove unnecessary reprocessing for mobile execution. Its average
encoder input length is only 3.67s on S25, reducing encoder execution to
161.6 ms and decoder execution to 37.1 ms. This allows \sysname to complete each
CPU-only streaming update within the inference interval even on the mobile
CPU. The X Plus results show the same trend under a stronger CPU: all systems
become less constrained, but \sysname still uses substantially shorter encoder
inputs and lower encoder/decoder latency than the baselines.

\textbf{Scope.} Note that these results are not intended as the main system comparison, but as a supplementary evaluation showing that on-device AI systems are highly sensitive to redundant computation. Eliminating those redundant computations for encoder and decoder is essential for sustaining real-time streaming on mobile processors. Jointly pipelining the CPU, GPU, and NPU remains an interesting direction for future work.

\section{GPU Implementation Details}\label{app:experiment-detail}

\textbf{Implementation and GPU setup.} All GPU implementations were built on \texttt{whisper.cpp} \citep{whispercpp}, with GGML used as the tensor computation backend. For GPU-accelerated execution, we used the GGML OpenCL backend on Qualcomm Adreno GPUs. Experiments were conducted on two devices: a Samsung Galaxy S25~\citep{samsung_galaxy_s25} equipped with a Qualcomm Adreno 830 GPU, and an ASUS Vivobook S 15~\cite{qualcomm_laptop_products} powered by Snapdragon X Plus with a Qualcomm Adreno X1-45 GPU. Both platforms used OpenCL 3.0 and were compiled in release mode with Adreno-optimized OpenCL kernels enabled. Unless otherwise stated, all experiments used the \texttt{ggml-base.bin} Whisper model, and streaming inference was performed with a 2 s inference round interval. The motivation experiments in \S~\ref{sec:motivation} were conducted on the Snapdragon X Plus using 200 samples from the LibriSpeech dataset~\citep{panayotov2015librispeech}.

\section{NPU Implementation Details}\label{app:implementation-detail}

\subsection{Graph Construction}
\label{app:npu-graph-lowering}

\textbf{ONNX/QNN setup.} For NPU execution, we use ONNX graphs executed with the ONNX Runtime QNN execution provider and the QAIRT/QNN HTP backend. The NPU implementation follows the QAI Hub Whisper graph construction: transformer linear layers are represented as 1$\times$1 convolution operators for QNN execution, while core Whisper operations such as attention, layer normalization, GELU, and encoder front-end convolutions are preserved. 

\begin{center}
\centering
\small
\setlength{\tabcolsep}{4pt}
\captionof{table}{Effect of QAI Hub linear-to-1$\times$1-convolution lowering on S25 NPU latency. Both variants use the same ONNX Runtime QNN/QAIRT stack and ONNX I/O contract.}
\label{tab:npu-graph-lowering}
\begin{tabular}{lrrr}
\toprule
\textbf{Metric} & \textbf{QAI Hub-lowered} & \textbf{HF-direct Linear} & \textbf{Reduction} \\
\midrule
Encoder \texttt{Session::Run} & 47.618 ms & 51.195 ms & 7.0\% \\
Decoder \texttt{Session::Run} sum & 139.417 ms & 158.354 ms & 12.0\% \\
Decoder total & 184.570 ms & 205.664 ms & 10.3\% \\
Preprocess + encoder + decoder & 235.060 ms & 259.955 ms & 9.6\% \\
\bottomrule
\end{tabular}
\end{center}

\textbf{Lowering effect.}
This lowering improves S25 NPU latency under the same runtime stack, so we use the lowered QNN graphs for our NPU implementation.

\subsection{Input and Model Bucketing}
\label{app:model-bucketing}

\textbf{Bucket selection.}
The NPU runtime uses input and model bucketing to connect variable-length
streaming audio with the static-shape execution model of Qualcomm QNN. Rather
than compiling a separate graph for every possible window length or padding every
input to Whisper's 30s context, \sysname selects the smallest supported bucket
that can contain the current effective audio window. Given an effective audio
duration \(d\), the selected bucket is
\[
    \mathcal{B}=\{3,4,5,6,30\}\ \text{s}, \qquad
    b(d)=\min\{\,\beta \in \mathcal{B} \mid d \le \beta\,\}.
\]
The 3--6s buckets are sufficient for most audio-buffer lengths in our
experiments, while the 30s graph is kept as a rare fallback for unstable segments
that require reinference with the native Whisper context.

\subsection{Controlled Unrolling Decoder Graphs}
\label{app:controlled-unrolling-graphs}

\textbf{Chunked decoder execution.}
The decoder is executed as a sequence of small static graphs rather than as one
graph that always runs the full decoding loop. This keeps each NPU graph
shape-specialized while allowing the runtime to stop early when no further
tokens are needed. For each bucket, the decoder supports up to 30 output tokens. After each chunk graph, the system emits valid tokens and dispatches the next graph only if decoding must continue.

\textbf{Chunk schedule and prompt prefill.}
For each bucket, the decoder is implemented as a controlled-unrolling chain with
a maximum decode length of 30 tokens. The default schedule is
[4,6,5,5,5,5], where the first chunk handles Whisper's four special
tokens. When carryover text is used as a decoder prompt, we use the prefill
schedule [6,4,5,5,5,5] so that prompt and special tokens are handled
within the first two chunks. Since both schedules cover 10 positions before the
later chunks, the remaining [5,5,5,5] graphs are shared across the
default and prefill paths. Each chunk returns logits, cross-attention maps, and
updated self-KV cache tensors, and chunk wrappers within each bucket are linked
into a weight-shared QNN context binary to minimize decoder weights.

\textbf{Chunk-size selection.}
We select the repeated later chunk size while keeping the first two chunks fixed as [4,6] for the default path and [6,4] for the prompt-prefill path. We compare later chunk sizes of 3, 5, and 7,
corresponding to default schedules of [4,6,3,3,3,3,3,3],
[4,6,5,5,5,5], and [4,6,7,7,7], respectively. This comparison
captures the trade-off between graph-dispatch overhead and unused decoder
computation: smaller chunks require more graph launches, while larger chunks
execute more decoder positions than may be needed.

\textbf{Selected chunk size.}
Table~\ref{tab:decoder-chunk-size-selection} shows that 3-token chunks
incur excessive decoder calls and graph switches, while 7-token chunks
perform more unnecessary decoder-position computation. The 5-token setting
best balances graph-dispatch overhead and unused decoder work, achieving the
lowest per-word latency. We therefore use [4,6,5,5,5,5] for the default
path and [6,4,5,5,5,5] for the prompt-prefill path.

\begin{table}[h]
\centering
\small
\caption{Chunk-size selection with fixed first two chunks. Per-word latency is measured as the difference between a word emission time and the end time of its matched ground-truth word. Extra work/token is a proxy for unused decoder computation, measured as the number of decoder token positions executed per emitted transcript token.}
\label{tab:decoder-chunk-size-selection}
\begin{tabular}{lcccc}
\toprule
\textbf{Chunk size}
& \textbf{Per-word latency (s)}
& \textbf{Calls/token}
& \textbf{Switches/token}
& \textbf{Extra work/token} \\
\midrule
3 & 1.994 & 0.452 & 0.575 & 1.827 \\
5 & 1.943 & 0.389 & 0.510 & 1.946 \\
7 & 2.015 & 0.356 & 0.475 & 2.040 \\
\bottomrule
\end{tabular}
\end{table}

\subsection{Cross-Attention Masking}
\label{sec:npu-cross-attention}

\textbf{Padded source positions.}
In the bucketed NPU runtime, the number of encoder frames attended by the decoder is fixed by the selected bucket rather than by the exact audio duration. When the streaming window is shorter than the bucket, the encoder output contains real audio positions followed by padded positions introduced only to match the static bucket shape. If left unmasked, these padded positions would still participate in the cross-attention softmax and could change the attention distribution over real audio frames.

\textbf{Masking padded positions.}
Each decoder chunk receives a cross-attention mask constructed from the non-padded audio length recorded during bucket selection. The mask leaves the columns corresponding to real audio unchanged and adds a finite negative bias of \(-100\) to the padded columns before the softmax. We use a finite value instead of infinity to avoid numerical issues under reduced-precision NPU execution. The final-layer cross-attention maps returned to the host are also cropped to the real-audio region before hallucination detection and DTW-based carryover updates. As a result, the compiled NPU graph keeps a fixed bucket shape, while token validation and streaming decisions are computed only from real audio.

\begin{figure}[h]
    \centering
    \includegraphics[width=0.95\linewidth]{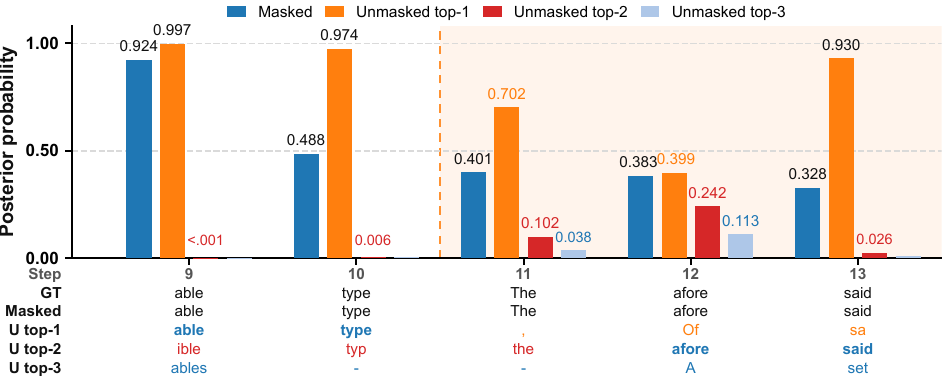}
    \caption{Effect of removing the cross-attention source-axis mask on
    LJ001-0021, decoder steps 9--13. Bars show the masked top-1 posterior and
    the unmasked top-1, top-2, and top-3 posteriors. At step 11, the masked token
    is absent from the unmasked top-3 candidates; at steps 12--13, it appears
    only as the second-ranked unmasked candidate.}
    \label{fig:cross-attn-mask-drift}
\end{figure}

\textbf{Mask-removal case study.}
Figure~\ref{fig:cross-attn-mask-drift} shows the token-level effect of removing
the cross-attention mask on the LJ-Speech utterance LJ001-0021~\citep{ljspeech17}. With the same
encoder input, decoder weights, and decoding state, the unmasked decoder leaves
the masked top-1 trajectory after step 10. At the first failure point, the masked
token is absent from the unmasked top-3 candidates; in the following steps, it is
demoted below an incorrect unmasked top-1 token.

\subsection{Self-Attention Cache Handling}
\label{sec:npu-self-attention}

\textbf{Static cache interface.}
Controlled unrolling executes multiple autoregressive decoding steps inside one
NPU graph, but the first chunk would normally start with an empty self-KV cache.
This zero-length cache is inconvenient for QNN compilation and would make the
first chunk graph a special case. We avoid this by reserving one leading dummy KV
slot, so every chunk starts with a non-empty cache and follows the same
read--append--return cache interface.

\textbf{Dummy-slot masking.}
The dummy slot is only a structural placeholder. It is masked as an invalid
self-attention source position, and the host never treats it as a generated
token. Thus, the cache interface remains static for NPU execution while token
emission, EOS checks, hallucination detection, and prompt carryover operate only
on real generated tokens.

\subsection{Software Stack}
\label{app:software-versions}

\textbf{NPU runtime stack.}
We implemented the bucketed NPU runtime using ONNX Runtime with Qualcomm's QNN
HTP backend. Encoder and decoder graphs were exported to ONNX and compiled
offline with Qualcomm AI Runtime SDK (QAIRT) \texttt{2.37.1.250807}. The same runtime, NPU backend, and offline
compiler stack were used across both target devices. The target-specific details
are the application build environments: Visual Studio 2022 ARM64/MSVC for the
Snapdragon X Plus laptop and Android NDK r26d with clang 17 for the Galaxy S25.

\begin{table}[h]
\centering
\caption{Shared NPU software stack.}
\label{tab:implementation_stack_shared}
\footnotesize
\setlength{\tabcolsep}{5pt}
\renewcommand{\arraystretch}{0.95}
\begin{tabular}{ll}
\toprule
\textbf{Item} & \textbf{Configuration} \\
\midrule
Runtime & ONNX Runtime QNN \texttt{1.23.2} \\
NPU backend & QAIRT/QNN HTP \texttt{2.37.1.250807} \\
Offline compiler & \texttt{qairt-converter}, \texttt{qnn-context-binary-generator} \\
Targets & Snapdragon X Plus laptop; Samsung Galaxy S25 \\
Builds & VS2022 ARM64/MSVC; Android NDK r26d, clang 17, \texttt{android-33} \\
\bottomrule
\end{tabular}
\end{table}